\documentclass[aps,reprint,superscriptaddress,longbibliography]{revtex4-1}

\usepackage{graphicx}

\newcommand{\DTSA}{\Delta T_\mathrm{SA}}

\newcommand{\Nu}{\mathrm{Nu}}\newcommand{\RaSA}{\mathrm{Ra}_\mathrm{SA}}

\newcommand{\vect}[1]{\mathbf{#1}}
\newcommand{\degC}{$^\circ$C}

\begin{document}


\title{Experimental study of convection in the compressible regime}



\author{R\'emi Menaut}
\affiliation{Universit\'e de Lyon, UCBL, ENSL, CNRS, LGL-TPE, 69622 Villeurbanne, France}

\author{Yoann Corre}
\affiliation{Universit\'e de Lyon, UCBL, ENSL, CNRS, LGL-TPE, 69622 Villeurbanne, France}

\author{Ludovic Huguet}
\affiliation{Aix-Marseille Universit\'e, CNRS, Centrale Marseille, IRPHE UMR 7342, Marseille, France}

\author{Thomas Le~Reun}
\affiliation{Aix-Marseille Universit\'e, CNRS, Centrale Marseille, IRPHE UMR 7342, Marseille, France}

\author{Thierry Alboussi\`ere}
\affiliation{Universit\'e de Lyon, UCBL, ENSL, CNRS, LGL-TPE, 69622 Villeurbanne, France}

\author{Michael Bergman}
\affiliation{Physics Department, Simon's Rock College, 84 Alford Road, Great Barrington, MA 01230, USA}

\author{Renaud Deguen}
\affiliation{Universit\'e de Lyon, UCBL, ENSL, CNRS, LGL-TPE, 69622 Villeurbanne, France}

\author{St\'ephane Labrosse}
\affiliation{Universit\'e de Lyon, UCBL, ENSL, CNRS, LGL-TPE, 69622 Villeurbanne, France}

\author{Marc Moulin}
\affiliation{Universit\'e de Lyon, ENSL, CNRS, Laboratoire de physique de l'ENSL, 69364 Lyon, France}




\date{\today}

\begin{abstract}
An experiment of thermal convection with significant compressible effects is presented. The high-gravity environment of a centrifuge and the choice of xenon gas enable us to observe an average adiabatic temperature gradient up to $3.5$~K~cm$^{-1}$ over a 4~cm high cavity. At the highest rotation rate investigated, 9990~rpm, the superadiabatic temperature difference applied to the gas layer is less than the adiabatic temperature difference. The convective regime is characterized by a large Rayleigh number, about $10^{12}$, and dominant Coriolis forces (Ekman number of order $10^{-6}$). The analysis of temperature and pressure fluctuations in our experiments shows that the dynamics of the flow is in a quasi-geostrophic regime. Still, a classical power law (exponent $0.3 \pm 0.04$) is observed between the Nusselt number (dimensionless heat flux) and the superadiabatic Rayleigh number (dimensionless superadiabatic temperature difference). However, a potential hysteresis is seen between this classical high flux regime and a lower heat flux regime. It is unclear whether this is due to compressible or Coriolis effects. In the transient regime of convection from an isothermal state, we observe a local decrease of temperature which can only be explained by adiabatic decompression.  

\end{abstract}


\maketitle

\section{Introduction}
Thermal convection is an important mechanism in the dynamical and thermal evolution of geophysical and astrophysical systems. One of the first theoretical description of this phenomenon was made by Boussinesq \cite{boussinesq1903} in an incompressible regime. This study was used by Rayleigh who obtained the criterion of stability of a layer of a fluid heated from below \cite{rayleigh1916}. 
First works on convection focused on incompressible fluids and could not be applied to geophysical and astrophysical systems in which compressibility is important due to large variations across large scale objects. To study these kinds of systems, \citet{ogura1961} proposed the anelastic approximation  in which acoustic waves vanish while other compressibility effects are retained. This approximation consists in considering convection as fluctuations around an isentropic state. The nearly hydrostatic pressure gradient and the isentropic hypothesis imply the existence of a temperature gradient, called the adiabatic gradient, from low temperatures at high altitude to high temperatures at low altitude, as first suggested by \citet{carnot}.

The anelastic approximation has been applied to many different natural objects such as the atmosphere \cite{ogura1961}, the Earth's outer core \cite{anufriev2005}, gas giant planets \cite{duarte2018} or stars \cite{elliott2008,jonesbenchmarks2011}. In addition, more theoretical studies have been conducted to gain a better understanding of phenomena which take place in compressible convection \cite{alboussiere2017,tilgner2011}. Nevertheless, all these studies are theoretical or based on numerical approaches. 

On the other side, many convection experiments have been done but the overwhelming majority of them are in the incompressible regime. In the geophysical and astrophysical fields, most of these experiments focus on the heat transfer due to convection by evaluating the Rayleigh-Nusselt relationship and on the influence of the rotation on this transfer. Different geometries (cylindrical cell, spherical, hemi-spherical) and fluids have been tested. A review of the obtained power laws was made by Aurnou \cite{aurnou2007}. 

The adiabatic gradient has been observed in the Earth's atmosphere by sounding balloons \cite{humphreys1909}, but there are very few experiments in which compressible convection effects are present. One study in a gas-pressurized Rayleigh-B\'enard cell \cite{ashkenazi1999} mentions the adiabatic gradient, and takes it into account to estimate a superadiabatic temperature difference. The total adiabatic temperature drop was $9.5$~mK (meaningful as the temperature control is within $0.4 $~mK) over a height equal to $105$~mm, using SF$_6$ close to the critical point. The maximum temperature difference applied was $10$~K, so that the adiabatic temperature difference introduced a small correction most of the time, except for the smallest temperature differences.


We present here an experiment especially designed to study compressible convection in the lab. The parameters of the experiment have been optimized for having significant compressible effects: we use xenon gas placed in a centrifuge. An important dimensionless parameter characterizing compressibility during convection is the dissipation number
\begin{equation} \label{Diss}
\mathcal{D} = \frac{\alpha g L}{c_p} ,
\end{equation}
where $\alpha$ and $c_p$ are typical values for the thermal expansion coefficient and heat capacity of the fluid, while $g$ and $L$ are typical values of gravity and size of the system. The dissipation number $\mathcal{D}$ is equal to the difference between the maximal (near the bottom) and minimal (near the top) values of the adiabatic temperature profile, divided by an average temperature in the system. The dissipation number is also close to the typical ratio of the viscous dissipation to the heat flux transferred across the system. Our experiments  
 reach a dissipation number of $0.06$,  still an order of magnitude less than that of the Earth's outer core or mantle. This value $\mathcal{D}=0.06$ is enough to reach an adiabatic profile of amplitude above 10~K. Usually, thermal convection experiments in a laboratory have a dissipation number below $10^{-5}$. 
In our experiment, we also have access to the departures of pressure and temperature away from their hydrostatic and adiabatic profiles. The values show that the pressure departures contribute significantly to entropy departures. This means that the anelastic liquid approximation should not be used to model our experiments, although the criterion for its validity seems to be met according to \citet{anufriev2005}. 

In a first part (section \ref{sec:adiabprofil}), we explain how the experiment was designed to obtain compressible effects in the laboratory and we derive expressions for the adiabatic profile of an ideal gas placed in a centrifuge. In section \ref{sec:exp}, we describe extensively our experimental setup and we study heat losses. In section \ref{sec:results}, we present the results of our experiment. We compare the temperature gradient to the adiabatic gradient. We then study the Rayleigh-Nusselt power law relationship. Next, we examine the temperature fluctuations and pressure signals to describe the flow dynamics. Finally, we study how convection is established in the initial transient of an experiment through the propagation of a convective front at the expense of a stably stratified state. Concluding remarks are made in section \ref{conclusion}.

\section{Adiabatic profile of xenon gas in a centrifuge}
\label{sec:adiabprofil}

In this section, we derive the analytic expression for the isentropic hydrostatic profile (also called the adiabatic profile) in the rotating frame of a rotor centrifuge for xenon gas. 
When convection is sufficiently developed in a compressible flow, the fluid state is isentropic due to the fast mixing of entropy, compared to the timescale of viscous or thermal dissipation. With the additional condition of hydrostatic equilibrium, a unique profile is obtained \cite{schwarzschild1906}. This profile, which is the neutral convective stability profile \cite{jeffreys1930}, is called the adiabatic profile. Thus the equations governing the adiabatic profile are
\begin{equation}
\nabla s_a = \vect{0},
\label{eq:gradadiab-s}
\end{equation}
\begin{equation}
\nabla p_a = \rho_a \vect{g},
\label{eq:gradadiab-P}
\end{equation}\begin{equation}
\nabla T_a = \alpha_s \vect{g},
\label{eq:gradadiab-T}
\end{equation}
where the subscript $a$ refers to the adiabatic profile, $s_a$ is the specific entropy, $T_a$ the temperature, $p_a$ the pressure, $\rho_a$ the density, $g$ the gravitational acceleration and $\alpha_s$ the isobaric entropy expansion coefficient
\begin{equation}
\alpha_s = -\frac{1}{\rho}\left(\frac{\partial \rho}{\partial s}\right)_{p} = \frac{\alpha T}{c_p},
\end{equation}
where $\alpha$ is the isobaric thermal expansion coefficient and $c_{p}$ the specific heat capacity at constant pressure.

The so-called adiabatic gradient is defined by Eq.~(\ref{eq:gradadiab-T}). It is easily deduced from Eq.~(\ref{eq:gradadiab-s}) and Eq.~(\ref{eq:gradadiab-P}) by using thermodynamic identities. In order to obtain sizeable effects of compressibility in the experiment, our goal was to maximize the adiabatic gradient. There are two ways to do so: artificially increase gravity $\vect{g}$ or choose a fluid with good thermodynamic properties (large $\alpha_s$ \textit{i.e.} large $\alpha T$ value and small $c_{p}$). We have been following both ways.

A convenient way to raise the value of gravity to a high level is to use a centrifuge which creates a radial acceleration of amplitude $r\Omega^2$ where $r$ is the distance from the rotation axis and $\Omega$ is the rotation rate. With a rotation rate of several thousands rotations per minute and a rotor size of several centimeters, the rotational acceleration can reach values several  thousands times larger than Earth's gravity. As a consequence, we will neglect Earth's gravity in the following and consider a purely radial gravity.

In order to maximize the adiabatic gradient, it is better to use a gas ($\alpha T \sim 1$) than a liquid ($\alpha T \sim 10^{-4}$). The best gas candidates are thus the ones with the smallest $c_{p}$, hence monoatomic gases with large molar masses. We decided to use xenon which is one of the monoatomic gas with the smaller  specific heat capacity due to its large molar mass. Radon gas is still better but was discarded because of its radioactivity. In the range of our experiments, $T\in[280\text{~K};330\text{~K}]$, $p\in[1.75\text{~MPa};2.25\text{~MPa}]$, xenon is not an ideal gas. Because there is no simple analytic equation of state for xenon in these conditions, we use the CoolProp library \cite{coolprop}  to evaluate the thermodynamic properties of xenon in the conditions of our experiments. This library uses the empirical data of xenon given by \citet{lemmon2006}. Thermo-physical properties of Xenon at 300~K and 2~Mbar are listed in Table \ref{tab:xenon}.

\begin{figure}
	\includegraphics{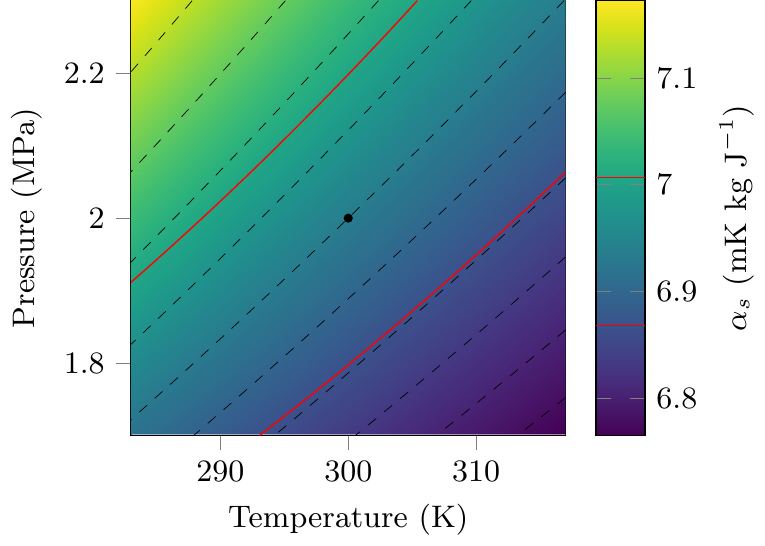}
	\caption{\label{fig:alphas} Value of $\alpha_s$ (from the CoolProp library) for our experiments. The dashed lines are the isentropic curves. The red lines delimit the area in which $\alpha_s$ variations are less than 1\% from its value at 300~K and 2~MPa.}
\end{figure}

On Fig.~\ref{fig:alphas}, we plot the isobaric entropy expansion coefficient $\alpha_s$ in the range of pressure and temperature in our experiments. We see that its value has very little variations, in particular along isentropic curves. For instance, red lines (isovalues of $\alpha_s$) are different but very close to adiabats (dashed lines). Hereafter, we take $\alpha_s$ constant, which is true to a very good approximation. With this approximation, the expressions for the temperature adiabatic profile is easily determined by solving Eq.~(\ref{eq:gradadiab-T})

\begin{equation}\label{eq:Tprofil}
T_a(r) = T_a^{max} + \frac{\alpha_s\Omega^2}{2} (r^2-r_{max}^2),
\end{equation}
where $T_a^{max}$ is the extrapolated temperature, on the adiabat, at the bottom plate at $r=r_{max}$.

This expression will be used in part \ref{sec:results} to compare the measured and adiabatic profiles. Any difference between temperature and the adiabatic profile will be called the superadiabatic temperature.

\begin{table}
\begin{tabular}{p{3.5cm}ccc}
\hline\hline
Molar mass & $M$& 0.1313&kg~mol$^{-1}$\\
Density & $\rho_0$ & 118 &kg~m$^{-3}$ \\
\hline
Specific heat capacity at constant volume& $c_v$ & 105 & J~K$^{-1}$~kg$^{-1}$ \\
Specific heat capacity at constant pressure& $c_p$ & 204 & J~K$^{-1}$~kg$^{-1}$ \\
Heat capacity ratio&$c_p/c_v$&1.98&\\
\hline
Isobaric thermal expansion coefficient&$\alpha$ & $4.73\times10^{-3}$&K$^{-1}$\\
Product $\alpha T$  &$\alpha T$ & 1.42&\\
Entropic thermal expansion coefficient& $\alpha_s$ & $6.94\times10^{-3}$&K~kg~J$^{-1}$\\
\hline
Thermal conductivity \cite{vargaftik1993}& $k$ & $6.54\times10^{-3}$&W~K$^{-1}$~m$^{-1}$ \\
Dynamic viscosity \cite{hanley1974}& $\eta$ & $2.46\times10^{-5}$&Pa~s \\
Prandtl number & $\eta c_p / k$& 0.77 & \\

\hline\hline
\end{tabular}
\caption{\label{tab:xenon} Xenon properties at 300~K and 2~MPa. Except for the thermal conductivity and the viscosity, all these data are evaluated with the CoolProp library \cite{coolprop} using the xenon equation of state given by \citet{lemmon2006}.}
\end{table}

\section{Experimental setup}
\label{sec:exp}

\subsection{Global description}
\begin{figure}
\includegraphics[width=8cm]{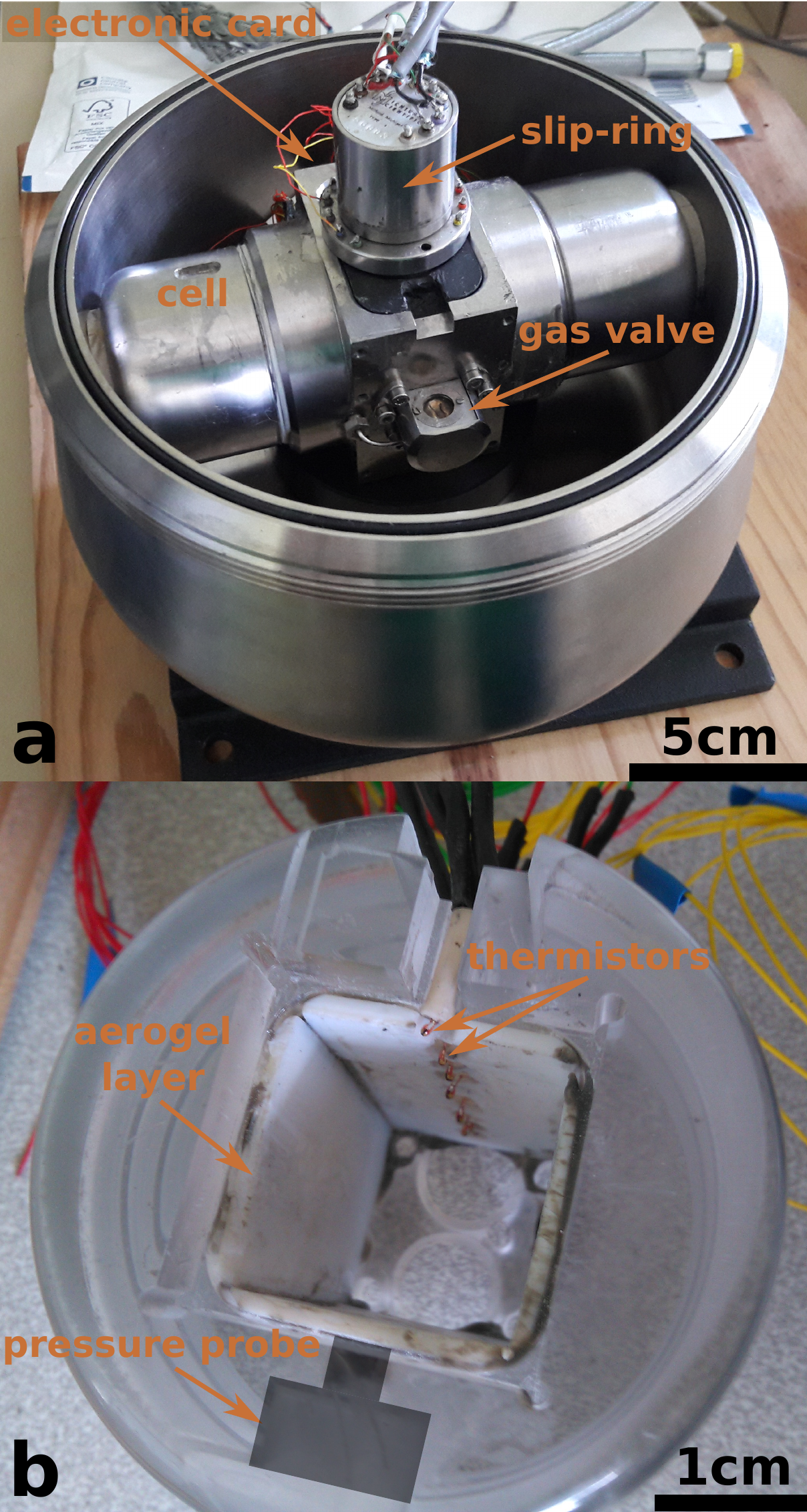}
\caption{\label{fig:photo} (a) Picture of the centrifuge rotor in which the cell is placed. (b) Polycarbonate cell with aerogel layer on the wall. The nine thermistors are the orange points aligned in the centre of a wall, see Fig.~\ref{fig:scheme}.}
\end{figure}

\begin{figure}
\includegraphics{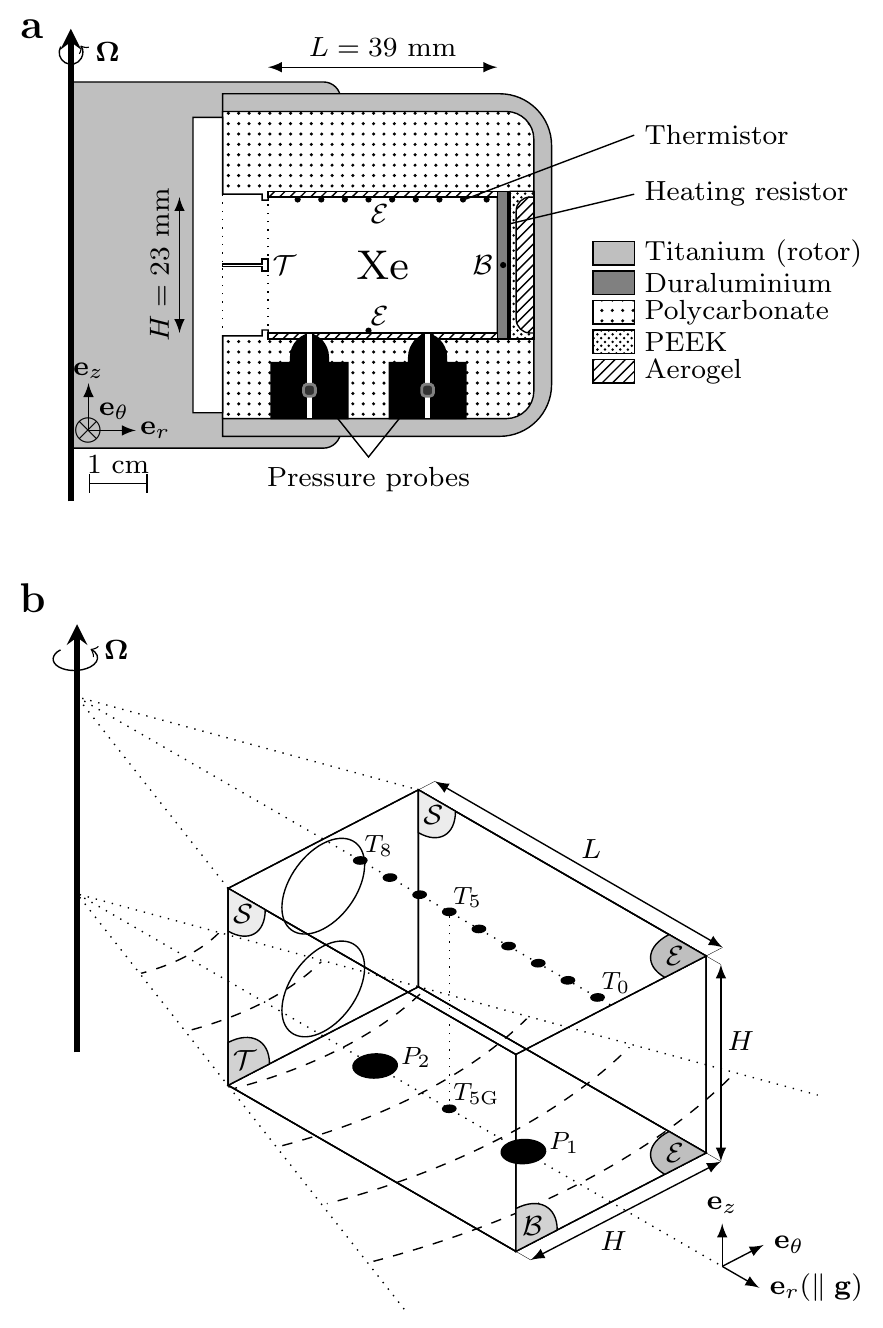}
\caption{\label{fig:scheme} (a) Sketch of the cavity of the experimental setup with heating resistor and thermal and pressure probe positions. (b) Three-dimension view of the rectangular cavity containing xenon gas. Small black circles are the thermistors positions. Big ones are the pressure probes positions. Dashed lines are the iso-gravity contour. Labels $\mathcal{B}$, $\mathcal{T}$, $\mathcal{E}$ and $\mathcal{S}$ correspond respectively to bottom wall, top wall, Ekman walls and Stewartson walls.}
\end{figure}

In our experiments, xenon gas is mostly contained in a cuboid cavity placed in a centrifuge (Figs.~\ref{fig:photo} and \ref{fig:scheme}). We use cylindrical coordinates $(\vect e_r, \vect e_\theta, \vect e_z)$. The rotation axis of the centrifuge $\vect e_z$ is vertical in the laboratory and the long axis of the cavity is along the radial direction $\vect e_r$. To simplify our explanations in the following, we give different names to the different walls of the cavity. The two walls of normal vector $\vect e_r$ are called the top and  bottom walls. The furthest from the rotation axis, where heat will be provided to the fluid, is the bottom wall and the closest to the axis, is the top wall. The other four faces are named according to the type of boundary layer that develops due to the rotation. Thus, walls of normal vector $\vect e_z$ are called Ekman walls and the last two walls are called Stewartson walls.

The cavity is cut into a polycarbonate cylinder fixed to the centrifuge's rotor by a titanium lid (Fig.~\ref{fig:photo}). The cavity is $L=39$~mm long with a section of $H^2=23\times 23$~mm$^{2}$. Ekman and Stewartson walls are thermally insulated by a 1~mm layer of aerogel (Airloy X103 Class M). On the top wall, there are two cylindrical holes with radii 6 mm  and 7.7 mm long through which xenon is in contact with the rotor's titanium. On the bottom wall, xenon is heated by a duraluminium  plate under which a square heating resistor is placed. This plate is held in place by a piece of PEEK plastic under which there is another layer of aerogel for thermal insulation. Properties of these materials are listed in Table \ref{tab:setup}. 

Temperatures are measured by nine NTC thermistors aligned in the centre of an Ekman wall. These thermistors are glass-encapsulated sensors with a head of 0.8~mm of diameter and a resistance of 10~k$\Omega$ at 25\degC. Moreover another thermistor is added on the opposite side. The hot and cold temperatures are measured using a thermistor placed in a hole at the centre of the duraluminium plate and another is fixed to the rotor. Two piezoelectric pressure probes allow us to measure dynamical pressure $p-p_a$. They measure the differential pressure between the cavity and the thin layer of gas, supposed to be at rest ({\it i.e.} without dynamical pressure), contained between the polycarbonate and the titanium lid.

\begin{table}
\begin{tabular}{cccc}\hline\hline
Materials & Thermal& Specific  & Density\\
 & conductivity & heat capacity & \\
&(W~m$^{-1}$~K$^{-1}$)&(J~K$^{-1}$~kg$^{-1}$)& (kg~m$^{-3}$)\\\hline
Aerogel & 0.029& 680 to 730 & 200\\
Polycarbonate & 0.19 to 0.22& 1200 to 1300 & 1200 \\
PEEK & 0.25& 320 & 1320\\
Duraluminium  & 134 & 920 & 2700\\
Titanium & 21.9& 522 & 4510\\ \hline\hline

\end{tabular}
\caption{\label{tab:setup} Thermal properties of materials in the setup.}
\end{table}

The connection with the probes inside the centrifuge during the experiments is made through a slip-ring (Michigan Scientific S10) with 10 channels. Due to the high number of signals to be measured, the signals are multiplexed using an electronic card fixed to the centrifuge's rotor. To simplify the electronic circuit, we only multiplex signals from the thermistors. We add a reference resistor of a known constant value and we link this resistor and thermistors together in a series circuit. The multiplexer is wired to return the voltage of each thermistor cyclically. There are 11 cycles of measurements per second. Because we use a 16 channels multiplexer, we measure the voltage of each thermistor every 91~ms on a window of 5.7~ms.   The principle of temperature measurement is that the intensity passing through the thermistors is the same. We determine the value of the intensity by measuring the voltage of the reference resistor, then we measure the voltage of each thermistor to evaluate their resistance. And finally, we deduce the corresponding temperature from an analytical expression representative of the relationship between temperature and resistance
\begin{equation}
R = R_0 e^{-\beta_0\left(\frac{1}{T_0}- \frac{1}{T}\right)},
\end{equation}
where $R$ is the resistance of the thermistor and $R_0 = 10$~k$\Omega$, $T_0=25 ^\circ$C, $\beta_0=3492$~K are given in the datasheet provided by the maker. The calibration of the thermistors is done in an isothermal environment, the value $R_0$ of each thermistor is slightly adjusted to match the imposed temperature. 

\subsection{Heat losses and heat capacity}
\label{ssec:heatlosses}

The setup is heated by the square heating resistor which dissipates a power $\Phi_T$ with a maximum value of 9.33~W. However, this power is not entirely transmitted to the xenon, a fraction of it is lost to the walls of the setup. Two quantities are important to describe this phenomenon, the setup's heat capacity $C$ which controls how the setup warms up during an experiment and its thermal resistance $R_{th}$ which quantifies the power lost by conduction $\Phi_{lost}$. The power effectively transmitted to the xenon $\Phi$, in the steady state, is written
\begin{equation}
\Phi = \Phi_T-\Phi_{lost},
\end{equation}
where
\begin{equation}
\Phi_{lost} = \frac{T_{hot}-T_{ext}}{R_{th}}.
\end{equation}
$T_{hot}$ is the temperature of the duraluminium plate, $T_{ext}$ is the temperature of titanium outside the cell.

\begin{figure}
\begin{center}
\includegraphics{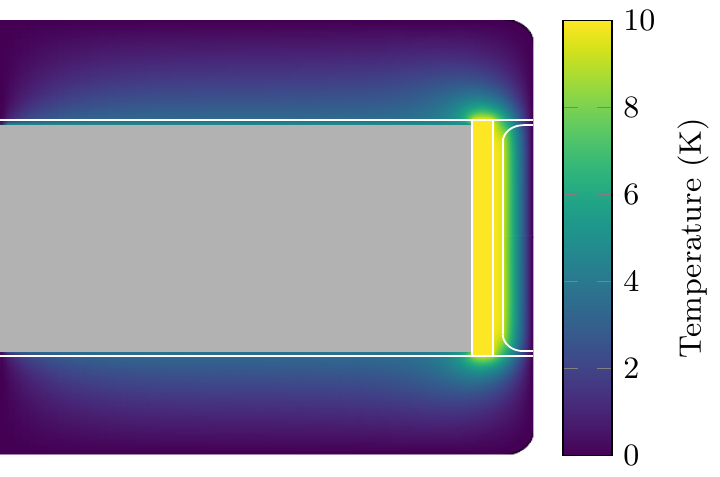}
\caption{\label{fig:thermal_simu} Thermal simulation of the setup. We impose a temperature difference of 10~K between the duraluminium plate and the exterior. The grey part is the xenon cavity assumed to be at the mean temperature between the duraluminium plate and the exterior.}
\end{center}
\end{figure}

To evaluate these two quantities, we made a thermal numerical simulation of the setup using the finite element solver FreeFem++ \cite{freefem}. Since the main part of the setup is cylindrical, we use an axisymmetric simulation and solve the corresponding 2D problem (see Fig.~\ref{fig:thermal_simu}). In our simulation, we replace the xenon cubic cavity by a cylindrical one such that the heating resistor and the duraluminium plate have the same surface area. In each simulation, we solve on the whole setup the heat equation
\begin{equation}
\rho_m c_{p,m} \frac{\partial T}{\partial t} = \nabla(k_m \nabla T),
\end{equation}
where $\rho_m$, $c_{p,m}$ and $k_m$ are the density, the specific heat capacity and the thermal conductivity of materials done in Table~\ref{tab:setup}. Moreover, we consider, as external boundary conditions, that the titanium lid stays at a constant temperature that we arbitrarily take as $T = 0$.

The first simulation is static ($\partial T / \partial t =0$) and determines the thermal resistance $R_{th}$. We imposed a known temperature difference $\Delta T = 10$~K between the duraluminium plate and the outside, and a uniform intermediate temperature $\Delta T/2$ in the xenon region. By calculating the heat flux $\Phi_{lost}$ which leaves the duraluminium plate through the faces not in contact with xenon, we estimate $R_{th}$ as
\begin{equation}
R_{th}= \frac{\Delta T}{\Phi_{lost}}.
\end{equation} We find a value of $R_{th}=\text{40~K~W$^{-1}$}$. However, this simulation cannot reproduce all the details of the setup. This is the reason for including an uncertainty of 25\% on this value in the following. Since $\Phi_T$ is well known, the uncertainty on $\Phi$ the effective power transmitted to xenon is caused by this uncertainty of 25\% on $\Phi_{lost}$. In this way, when the heating $\Phi_T$ is low, the uncertainty on $\Phi$ makes it difficult to know the effective flux transmitted to the xenon.

The second simulation determines the heat capacity $C$ of the walls of the setup. We start from an isothermal state at $T_0$ then, for $t>0$, we heat with a power $\Phi_T$ the area where the heating resistor is placed. In this simulation, we consider that the xenon is thermally inert \textit{i.e.} its heat capacity and thermal conductivity are zero. By considering the mean temperature $T_{hot}(t)$ of the duraluminium plate, we estimate $C$ as
\begin{equation}
C(t) = \frac{\Phi_T}{\frac{\partial T_{hot}}{\partial t}}.
\end{equation}
In the first 30~s, we find a value of $C$ which goes quasi-linearly from 5~J/K to 9~J/K. We do not need more precision here. Due to the complexity of the setup geometry, we are essentially looking for its order of magnitude.

\section{Results}\label{sec:results}

\subsection{Protocol}

\begin{figure}
\includegraphics{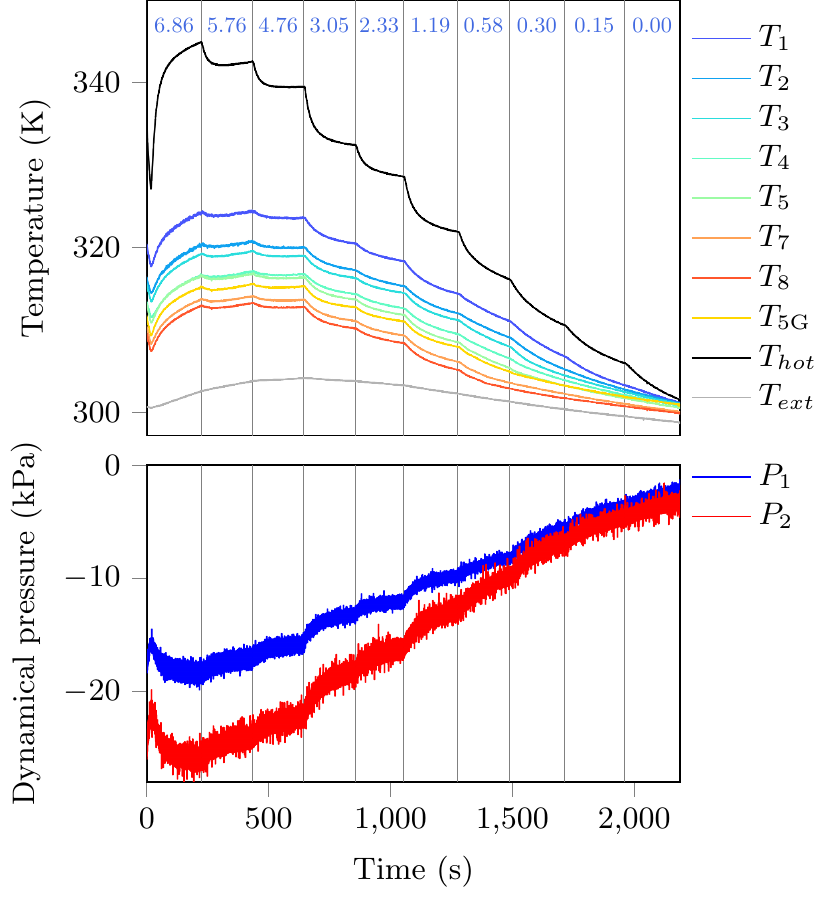}
\caption{\label{fig:profilT} Measured temperatures and pressures during an experiment at 9990 rpm. The blue numbers at the top indicate the heat power $\Phi_T$ in Watt. The gold line is the temperature $T_{5\text{G}}$ from the thermistor opposite $T_5$, the black line is the temperature of the heating plate $T_{hot}$ and the gray line is the titanium temperature outside the cell $T_{ext}$. $P_1$ corresponds to the pressure from the closest probe to the heating plate, $P_2$ to the furthest.}
\end{figure}

We ran experiments at three different rotation speeds 5000, 7000 and 9990 rpm. During an experiment, we set a constant rotation speed and we set different values of heat power in the heating resistor. For each value of the heat flux, we wait approximatively 3 minutes to reach an equilibrium state before changing its value. In order to avoid heating the whole setup above 80\degC~(to preserve the aerogel insulating material), we start with the highest heat power and decrease it step by step (Fig. \ref{fig:profilT}). The experiments done are listed in Table \ref{tab:experiment}.

\def\checkmark{$\bullet$}
\begin{table}[ht]
\begin{center}
\begin{tabular}{c|cccccccccccc}\hline\hline
$\Phi_T$ (Watt)&0&0.07&0.15&0.3&0.58&1.2&2.3&3&4.8&5.8&6.9&9.3\\\hline
5000 rpm&\checkmark&&&&&\checkmark&\checkmark&&\checkmark&&&\\
7000 rpm&\checkmark&\checkmark&\checkmark&\checkmark&\checkmark&\checkmark&\checkmark&&\checkmark&\checkmark&\checkmark&\checkmark\\
9990 rpm&\checkmark&\checkmark&\checkmark&\checkmark&\checkmark&\checkmark&\checkmark&\checkmark&\checkmark&\checkmark&\checkmark&\\ \hline\hline
\end{tabular}
\end{center}
\caption{Range of the experiments, in terms of rotation rate and heating power injected in the resistor.\label{tab:experiment}}
\end{table}

\subsection{Adiabatic gradient}
\label{ssec:AdiabaticGradient}

\begin{figure}
\includegraphics{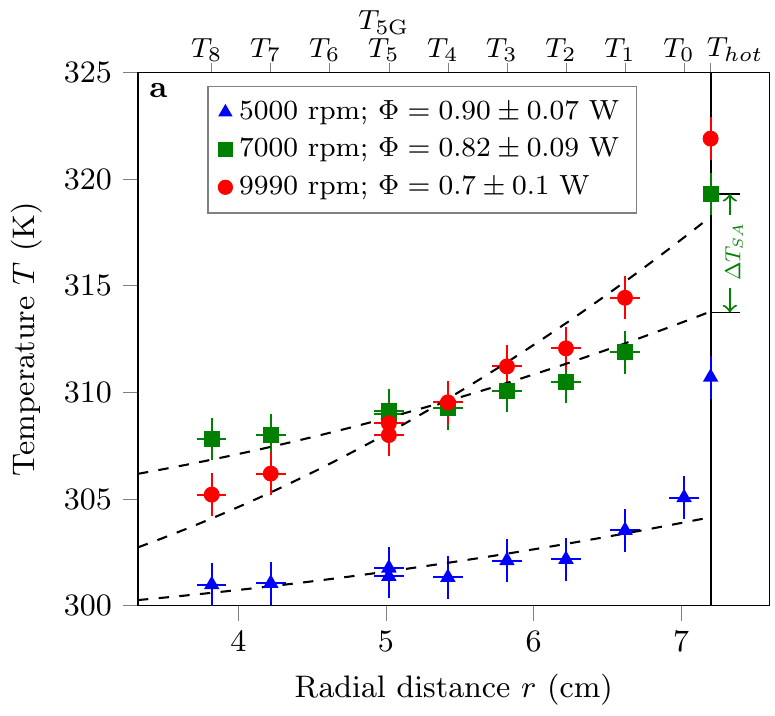}
\includegraphics{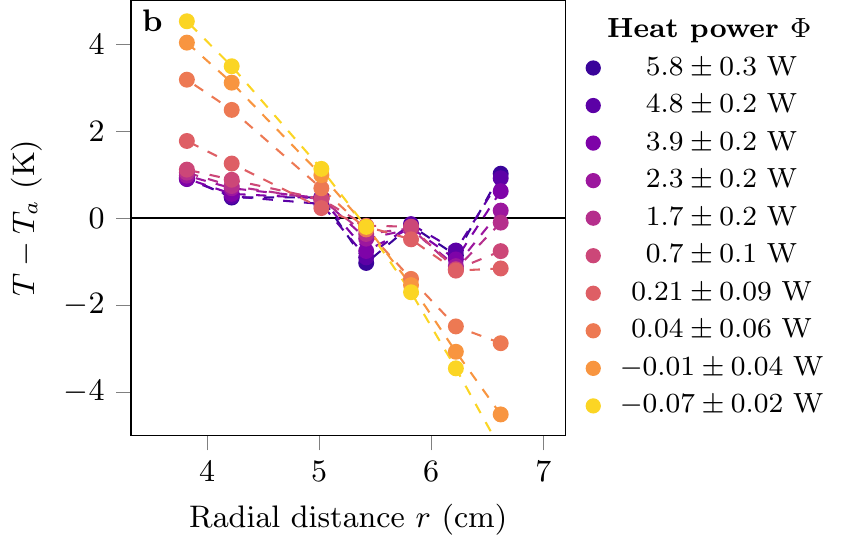}
\caption{\label{fig:gradadiab} (a) Temperature profiles averaged over $30$~s for $\Omega = 5000,7000,9990 \text{rpm}$ for an electric power dissipated in the heating resistor of $\Phi_T =$ 1.19~W. The dashed lines are theoretical profiles defined by Eq.~(\ref{eq:Tprofil}) where $T_a^{max}$ is adjusted by least-square method. The vertical black line is the bottom boundary of the xenon cavity. (b) Differences between temperature profile $T$ and theoretical profiles $T_a$ for different values of $\Phi$ at $\Omega = 9990 \text{~rpm}$.}
\end{figure}

When the heating is strong enough, we observe that the temperature profile in the cell is no longer isothermal. It is instead close to an adiabatic profile: by adjusting the reference temperature $T_0$ (not imposed in our setup) in Eq. (\ref{eq:Tprofil}) {\it via} a least square method on $T_1$ to $T_8$ (excluding $T_{hot}$ and $T_{ext}$), we obtain a very good agreement between the theoretical and experimental curves (Fig. \ref{fig:gradadiab}(a)). As expected, the profile shape depends only on gravity ({\it i.e.} rotation rate) outside of the boundary layer for high enough heating power. We measure an average adiabatic gradient of order 3.5~K/cm. 
On table \ref{tab:gravity}, we compute the maximum acceleration in the setup near the heating plate $r_{max} = 0.072$~m. From this value, we determine the maximum adiabatic temperature gradient and maximum value of the dissipation number defined in equation (\ref{Diss}), 
with the maximum gravity $g$, the height of the cavity $L=0.039$~m and an arbitrary representative value of the thermal expansion coefficient $\alpha = 4.73 \times 10^{-3}$~K$^{-1}$ (see Table~\ref{tab:xenon}). We also introduce another dimensionless number, the Ekman number $\mathrm{E}$, characterizing the magnitude of viscous forces compared to Coriolis apparent acceleration
\begin{equation}
\mathrm{E} = \frac{\nu}{\Omega H^2}, \label{Ekman}
\end{equation} 
where $\nu = \eta / \rho_0 $ is the kinematic viscosity, and the width of the cavity $H=23$~mm is used instead of its length $L=39$~mm. 

\begin{table}
\begin{tabular}{ccccc}\hline\hline
Rotation  & Apparent & Adiabatic  & Dissipation & Ekman \\
rate & gravity & gradient & number & number \\
 (rpm) & (m s$^{-2}$) & (K~m$^{-1}$)&&\\\hline
0 & 9.81 & 0.068 & 8.8~$\times$~$10^{-6}$ & $\infty$ \\
5000 & 19700 & 137 & 0.018 & 7.5~$\times$~$10^{-7}$ \\
7000 & 38700 & 268 & 0.035 & 5.4~$\times$~$10^{-7}$\\
9990 & 78800 & 547 & 0.071 & 3.8~$\times$~$10^{-7}$ \\ \hline\hline
\end{tabular}
\caption{\label{tab:gravity} Apparent gravity (centripetal acceleration) calculated near the heating plate ($r = 7.2$~cm), adiabatic gradient, dissipation number $\alpha g L / c_p$ and Ekman number $\nu/(\Omega H^2)$.}
\end{table}

When the heating is low, the temperature profile is not adiabatic but nearly isothermal. The threshold appears to be when the effective heating is close to zero. Actually, no adiabatic gradient is detected when 0~W is in the confident interval of heating. The uncertainty on the effective heating flux makes it difficult to estimate precisely this threshold.
On Fig.~\ref{fig:gradadiab}(b), we plot the difference between the temperature profile and the adiabatic profile from Eq.~\ref{eq:gradadiab-T} at 9990~rpm for various heat fluxes. It seems that there is an optimum heat flux corresponding to the best fit of the average temperature profile with the adiabatic profile. At $\Omega=9990$~rpm, this optimum flux is $\Phi  = 0.7$~W, (corresponding to the total flux $\Phi _T = 1.19$~W, on Fig.~\ref{fig:profilT}). At lower heat fluxes, the intensity of convection is probably not strong enough to impose an adiabatic profile (the curve with effective heat flux $-70 \pm 20$~mW is isothermal), while at higher fluxes the superadiabatic temperature contributions become large enough to alter the adiabatic profile. It can be seen on Fig.~\ref{fig:profilT} that the superadiabatic temperature difference becomes progressively larger than the total adiabatic temperature drop across the cavity when the flux increases above $\Phi _T = 1.19$~W. 

\begin{figure}
	\includegraphics{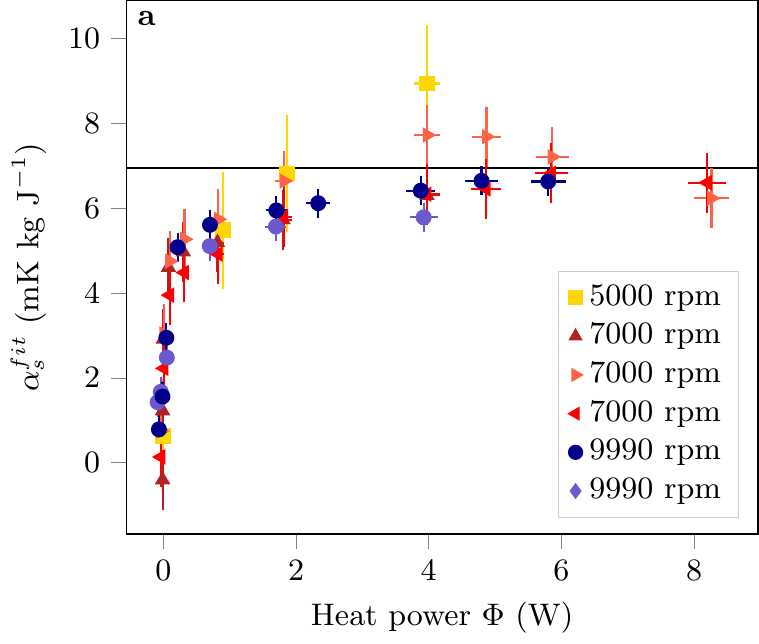}
	\includegraphics{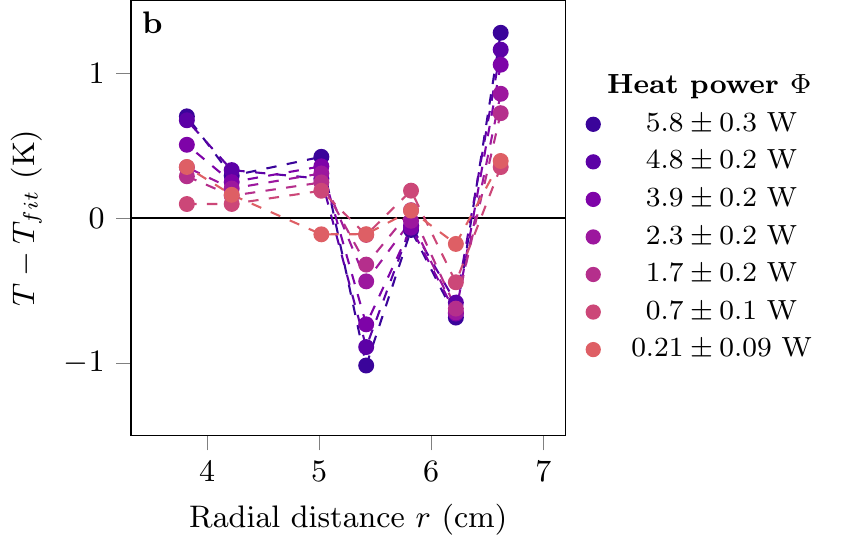}
	\caption{\label{fig:doublefit} (a) Best fitted quadratic coefficient of the temperature profile  $\alpha_s^{fit}$ from Eq.~(\ref{eq:doublefit}) as a function of the effective heat flux $\Phi$. The horizontal line corresponds to the adiabatic profile $\alpha_s^{fit}=\alpha_s=6.94\times 10^{-3}$~K~kg~J$^{-1}$. Each symbol color and shape corresponds to a series of measurements in which the heat flux is decreased from its highest to its lowest value. (b) Differences between temperature profile $T$ and fitted profiles $T_{fit}$ for different values of $\Phi$ at $\Omega = 9990 \text{~rpm}$.}
\end{figure}

Instead of using the theoretical value of $\alpha_s$, we have also tried to determine a value $\alpha_s^{fit}$ to fit the experimental temperatures data in the bulk of the fluid with a quadratic radial function of the form
\begin{equation}\label{eq:doublefit}
T_{fit}(r) = T_{fit}^{max} + \alpha_s^{fit} \frac{\Omega^2}{2}(r^2-r_{max}^2).
\end{equation}
Using a least square method, we determine the two parameters $T_{fit}^{max}$ and $\alpha_s^{fit}$. Values of $\alpha_s^{fit}$ are plotted on Fig.~\ref{fig:doublefit}(a) as a function of the effective heat flux for different rotation rates. For the smallest heat fluxes (tens of mW) the temperature profile is nearly isothermal corresponding to a vanishing $\alpha_s^{fit}$. Below a heat flux of 100~mW, the value of $\alpha_s^{fit}$ is significantly lower than the adiabatic value $\alpha_s$. Above 100~mW, increasing the heat flux brings $\alpha_s^{fit}$ closer to $\alpha_s$. On Fig.~\ref{fig:doublefit}(b), we plot the difference between the temperature profile and the expression (\ref{eq:doublefit}) using $\alpha_s^{fit}$ for values of heat fluxes above 200~mW. Fig.~\ref{fig:doublefit}(b) is analogous to Fig.~\ref{fig:gradadiab}(b) with reference to Eq.~(\ref{eq:doublefit}) instead of the adiabatic profile (\ref{eq:gradadiab-T}). We observe on  Fig.~\ref{fig:doublefit}(b) that the departure from the quadratic fit (\ref{eq:doublefit}) increases as the heat flux increases.

The temperature drop across the hot thermal boundary layer is estimated as follows: the quadratic curve fitting the experimentally measured temperatures is extrapolated at the radius of the hot surface, we then take the difference between the temperature measured within the hot plate and that extrapolation. Unfortunately, the boundary conditions are not well defined for the top cold plate due to the geometry of the setup. This is the reason why we will only consider the jump of temperature across the bottom hot boundary layer and call it the superadiabatic thermal difference $\DTSA$. It is defined, as shown on Fig.~\ref{fig:gradadiab}(a), by
\begin{equation}\label{dtsa}
\DTSA = T_{hot}-T_a^{max}.
\end{equation}
Experimentally, we find $\DTSA$ of order 10~K.

\subsection{Turbulent heat transfer}

From $\DTSA$ in equation (\ref{dtsa}), we wish to introduce a superadiabatic Rayleigh number $\RaSA$ and a Nusselt number $\Nu$. Because only the bottom boundary conditions are well established in our setup, it is natural to build these numbers with parameters pertaining to the bottom boundary layer. We define the superadiabatic Rayleigh number as
\begin{equation}
\RaSA = \frac{\rho_B^2 c_{p,B}^2 \alpha_{s,B} r_{max} \Omega^2 L^3 \DTSA}{T_B k \eta},
\end{equation}

where $T_B$, $\rho_B$, $c_{p,B}$ and $\alpha_{s,B}$ are the characteristic temperature, density, specific heat capacity and entropy expansion coefficient inside the bottom boundary layer. These quantities are evaluated numerically via the following algorithm.

Experimentally, we measure $T_a^{max}$ and $\DTSA$ with a best fit of the profile (\ref{eq:Tprofil}). Using the CoolProp library, we solve numerically Eq.~\ref{eq:gradadiab-T} and Eq.~\ref{eq:gradadiab-P} with the conditions that $T_a(r_{max})=T_a^{max}$ and the conservation of the entire xenon's mass. This results in the adiabatic pressure and density profiles $P_a(r)$ and $\rho_a(r)$ inside the cell. We note  $P_a^{max}$ and $\rho_a^{max}$ their values at $r=r_{max}$. Due to the small thickness of the boundary layer, we assume that the pressure stays constant across it. The characteristic pressure inside the boundary layer is thus $P_B=P_a^{max}$. The characteristic temperature inside the boundary layer $T_B$ is estimated as $T_B=T_a^{max}+\frac{1}{2}\DTSA$. Finally, from $T_B$ and $P_B$, the CoolProp library allows us to estimate all other quantities in the boundary layer $\rho_B$, $c_{p,B}$ and $\alpha_{s,B}$. Table \ref{tab:boundary} gives estimates of these parameters for different values of $\Omega$, $T_a^{max}$ and $\DTSA$.

\begin{table}
	
\begin{tabular}{c||c|ccccccc}\hline\hline\multicolumn{9}{c}{$\mathbf{\Omega = 5000}$~\textbf{rpm}}    \\\hline\hline
	&$T_a^{max}$&300&305&310&315&320&325&330\\
	&$\rho_a^{max}$&120&120&120&120&120&120&120\\
	$\DTSA$ &$P_a^{max}=P_B$&2.02&2.06&2.10&2.14&2.18&2.23&2.27\\
	\hline\hline5&$T_B$&302.5&307.5&312.5&317.5&322.5&327.5&332.5\\
	&$\rho_B$&118&118&118&118&118&118&118\\
	&$c_{p,B}$&203&202&200&198&197&196&194\\
	&$\alpha_{s,B}$&6.93&6.92&6.90&6.89&6.87&6.86&6.85\\
	\hline
	10&$T_B$&305.0&310.0&315.0&320.0&325.0&330.0&335.0\\
	&$\rho_B$&117&117&117&117&117&117&117\\
	&$c_{p,B}$&202&200&199&197&196&194&193\\
	&$\alpha_{s,B}$&6.92&6.90&6.89&6.88&6.86&6.85&6.84\\
	\hline
	15&$T_B$&307.5&312.5&317.5&322.5&327.5&332.5&337.5\\
	&$\rho_B$&116&116&116&116&116&116&116\\
	&$c_{p,B}$&200&199&197&196&195&193&192\\
	&$\alpha_{s,B}$&6.90&6.89&6.88&6.86&6.85&6.84&6.83\\
	\hline
	\hline\multicolumn{9}{c}{$\mathbf{\Omega = 7000}$~\textbf{rpm}}    \\\hline\hline
	&$T_a^{max}$&300&305&310&315&320&325&330\\
	&$\rho_a^{max}$&121&121&121&121&121&121&121\\
	$\DTSA$ &$P_a^{max}=P_B$&2.04&2.08&2.12&2.16&2.20&2.24&2.29\\
	\hline\hline5&$T_B$&302.5&307.5&312.5&317.5&322.5&327.5&332.5\\
	&$\rho_B$&120&119&119&119&119&119&119\\
	&$c_{p,B}$&204&202&200&199&197&196&195\\
	&$\alpha_{s,B}$&6.94&6.92&6.91&6.89&6.88&6.87&6.85\\
	\hline
	10&$T_B$&305.0&310.0&315.0&320.0&325.0&330.0&335.0\\
	&$\rho_B$&118&118&118&118&118&118&118\\
	&$c_{p,B}$&202&201&199&198&196&195&194\\
	&$\alpha_{s,B}$&6.92&6.91&6.89&6.88&6.87&6.85&6.84\\
	\hline
	15&$T_B$&307.5&312.5&317.5&322.5&327.5&332.5&337.5\\
	&$\rho_B$&117&117&117&117&117&117&117\\
	&$c_{p,B}$&201&199&198&196&195&194&193\\
	&$\alpha_{s,B}$&6.91&6.89&6.88&6.87&6.86&6.84&6.83\\
	\hline
	\hline\multicolumn{9}{c}{$\mathbf{\Omega = 9990}$~\textbf{rpm}}    \\\hline\hline
	&$T_a^{max}$&300&305&310&315&320&325&330\\
	&$\rho_a^{max}$&124&124&123&123&123&123&123\\
	$\DTSA$ &$P_a^{max}=P_B$&2.08&2.12&2.16&2.20&2.24&2.29&2.33\\
	\hline\hline5&$T_B$&302.5&307.5&312.5&317.5&322.5&327.5&332.5\\
	&$\rho_B$&122&122&122&122&122&122&122\\
	&$c_{p,B}$&205&203&201&200&198&197&195\\
	&$\alpha_{s,B}$&6.95&6.93&6.92&6.91&6.89&6.88&6.86\\
	\hline
	10&$T_B$&305.0&310.0&315.0&320.0&325.0&330.0&335.0\\
	&$\rho_B$&121&121&121&121&121&121&121\\
	&$c_{p,B}$&204&202&200&199&197&196&194\\
	&$\alpha_{s,B}$&6.94&6.92&6.91&6.89&6.88&6.87&6.85\\
	\hline
	15&$T_B$&307.5&312.5&317.5&322.5&327.5&332.5&337.5\\
	&$\rho_B$&119&119&119&119&119&119&119\\
	&$c_{p,B}$&202&200&199&197&196&195&193\\
	&$\alpha_{s,B}$&6.92&6.91&6.89&6.88&6.87&6.85&6.84\\
	\hline\hline\end{tabular}

	\caption{\label{tab:boundary} Calculated parameters in the boundary layer for different values of $\Omega$, $T_a^{max}$ and $\DTSA$. The units for temperature, pressure, density, heat capacity and entropy expansion coefficient are K, MPa, kg~m$^{-3}$, J~kg$^{-1}$~K$^{-1}$ and mK~kg~J$^{-1}$ respectively.}
\end{table}

The Nusselt number is defined as
\begin{equation}
\Nu = \frac{\Phi L }{k \DTSA S}.
\end{equation}
where $S = H^2 = 530$~mm$^2$ is the cross-sectional area. We could have defined the Nusselt with $\Phi-\Phi_a$ instead of $\Phi$ where $\Phi_a$ is the power conducted along the adiabat. This conduction heat flux is evaluated as $\Phi_a = k \nabla T_a S \sim \frac{kr_0 \Omega^2}{c_p} S \sim \text{10$^{-3}$~W}$. This value is smaller than typical values of $\Phi$ by 3 orders of magnitude, so we neglect it. An important point is that these estimates of Rayleigh and Nusselt numbers are valid only when the convection is established because it uses the adiabatic temperature fit (\ref{eq:Tprofil}) to obtain $\DTSA$.

\begin{figure}
	\includegraphics{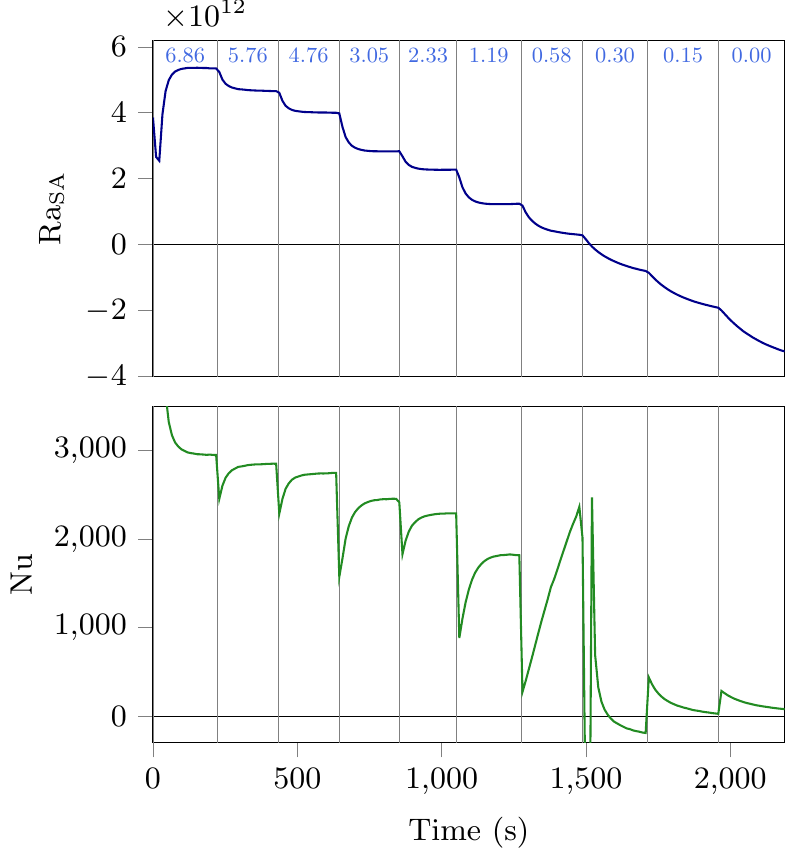}
	\caption{\label{fig:RaNuTime} Evolution of superadiabatic Rayleigh number $\RaSA$ and Nusselt number $\Nu$ during the same experiment as in Fig.~\ref{fig:profilT} at 9990~rpm. The blue numbers at the top indicate the heat power $\Phi_T$ in Watt.}
\end{figure}

\begin{figure}
\includegraphics{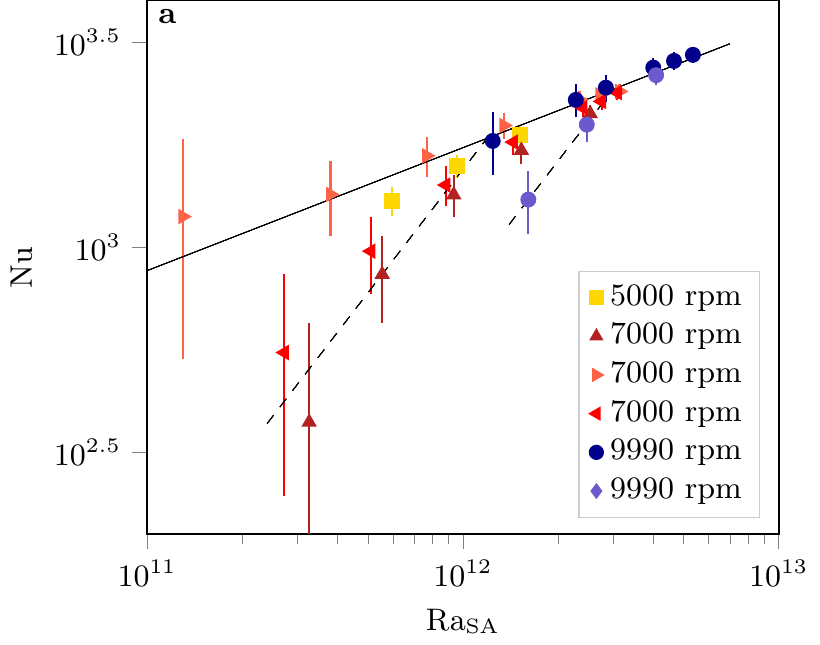}
\includegraphics{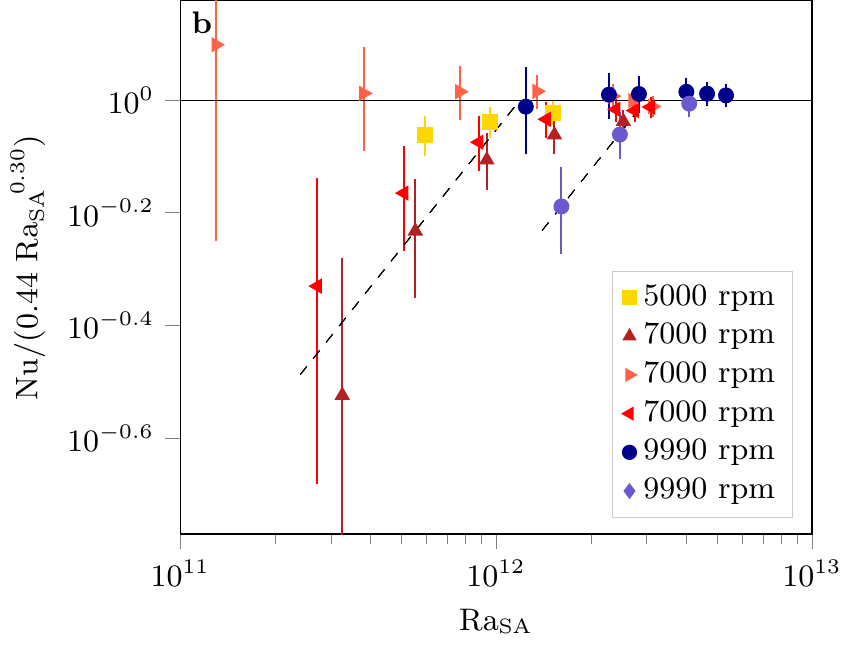}
\caption{\label{fig:RaNu} (a) Nusselt number values plotted as a function of the Rayleigh number, for different rotation rates. Each symbol color and shape corresponds to a series of measurements in which the heat flux is decreased from its highest to its lowest value. The solid line has a slope of $0.3$ and the dashed line a slope of $1$. (b) Same plot using the normalizing factor $0.44\, \mathrm{Ra}_{\text{SA}}^{0.30}$ for the Nusselt number.}
\end{figure}

We show on Fig.~\ref{fig:RaNuTime} the evolution of $\RaSA$ and $\Nu$ with time. When the heat flux is changed, the system reaches a new steady state where $\RaSA$ and $\Nu$ become constant in approximatively 1~min. We shall use those steady-state values to plot the Nusselt number in terms of the superadiabatic Rayleigh number on Fig.~\ref{fig:RaNu}. On Fig.~\ref{fig:RaNuTime}, for the total heat flux $\Phi _T = 0.58$~W and below, we are in the range where the net heat flux $Phi$ entering the gas volume is so weak that the adiabatic profile is not maintained in the cavity. Hence, the determination of the superadiabatic temperature difference becomes negative, which is meaningless. So, from $\Phi _T = 0.58$~W and below, both the superadiabatic Rayleigh number and the Nusselt number are not properly defined, because $\Delta T _{SA}$ is not properly defined and because the net heat flux $\Phi$ is not well known. We can nevertheless notice that these ill-defined quantities are governed by slow diffusive processes (heat conduction) since they do not reach a steady value within three minutes. 

Nusselt numbers obtained as described above are plotted as a function of the superadiabatic Rayleigh number on Fig.~\ref{fig:RaNu}. 
In order to model the $\Nu$-$\RaSA$ relationship, we look for a power law of the form
 \begin{equation}
\Nu \propto (\RaSA)^\beta.
\end{equation}
Our data are mostly consistent with a 1/3 power law (Fig. \ref{fig:RaNu}). Nevertheless, some data do not follow the same power law. When we run an experiment, we decrease the heating, and so $\RaSA$, step by step. Sometimes for low $\RaSA$, the heat transfer seems to follow another branch with a steeper power law and lower values. If we consider only points on the main branch we find the power law
\begin{equation}
\Nu = (0.44\pm0.02) \times (\RaSA)^{0.30\pm 0.04}. 
\end{equation}

The fact that some points are not on the main branch is surprising because, by reproducing the experiment at the same rotation speed, we observe sometimes the classic branch (exponent $0.3 \pm 0.04$) and some other times the steeper branch (exponent around $1$). It may be the signature of an hysteresis in the system. \citet{guervilly2016} showed that hysteresis is possible in rotating convection in a sphere in a quasi-geostrophic approximation. The geometry is not the same here but we expect that a similar phenomenon could explain this behavior.

Here we would like to draw the attention to what has been observed in another context, where a dual power law has been seen, while we think the underlying physics has no relationship with our experiments. In the case when rotation axis and gravity are parallel, two exponents have been found for the $\mathrm{Ra}$-$\Nu$ relationship \cite{king2009}, a steep law $\Nu\sim \mathrm{Ra}^{6/5}$ followed by the non-rotating usual law $\Nu\sim \mathrm{Ra}^{0.3}$. This is understood as the transition between a regime where the flow structures extend over the whole height of the cavity for moderate $\mathrm{Ra}$, followed by a 3D (yet anisotropic) dynamics at large $\mathrm{Ra}$. However, there is no hysteresis in this configuration, the heat transfer is determined by the value of the input parameters Rayleigh and Ekman numbers. On the contrary, our experiments show some hysteresis between two quasi-geostrophic configurations of seemingly different heat transfer power laws.

\subsection{Temperature measurements}

In fully convective cases, temperature signals can be seen as the sum of three contributions: the adiabatic profile, a stationary deviation from the adiabatic profile and finally, temporal fluctuations 
\begin{equation}
T = T_a + T_s + T'.
\end{equation}
At large rotation rate and heat flux, a typical amplitude of these contributions is 310~K for $T_a$ (with a 10~K increase from the top to the bottom), 1~K for $T_s$ and 0.05~K for $T'$. The profile of $T_a$ is given in Eq.~(\ref{eq:Tprofil}) and $T_s$ can be seen on Fig.~\ref{fig:gradadiab}(b) where the adiabatic profile has been subtracted to the time averaged temperature profile. The stationary departure $T_s$ from the adiabatic profile is interpreted as the signature of a stationary convective flow contribution.

\begin{figure}
\includegraphics{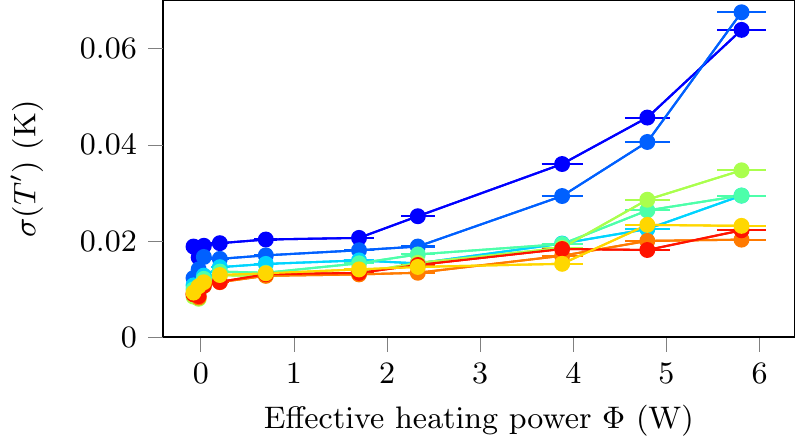}
\caption{\label{fig:TempSTD} Standard deviation of temperature fluctuations $\sigma(T')$ as a function of the heating power $\Phi$ at 9990 rpm. The color code is the same as in Fig.~\ref{fig:profilT}.}
\end{figure}

\begin{figure}
\includegraphics{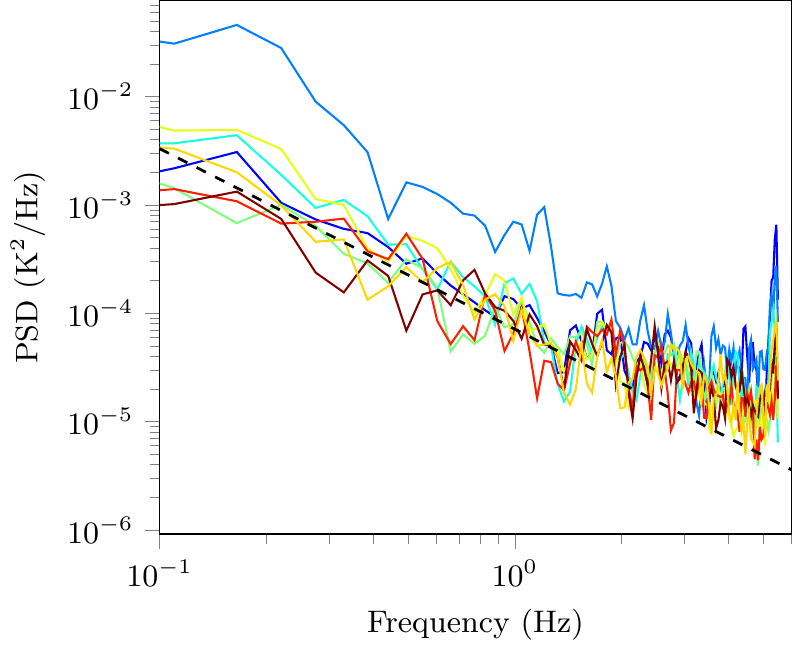}
\caption{\label{fig:temperature-spectrum} Power spectral density of temperature fluctuations at 7000~rpm and $\Phi=4.0\pm0.2$~W. The color code of temperature signals is the same as in figure \ref{fig:profilT}. The black dashed line shows a -5/3 slope.}
\end{figure}

We have access to temperature fluctuations by eliminating the long-term variations of temperature signals below 0.2 Hz. However, the study of temperature fluctuations is difficult  due to the high level of electronic noise on the signals caused by the multiplexing electronic card and the slip-ring. We show on Fig.~\ref{fig:TempSTD} the standard deviation of temperature fluctuations as a function of the heating power. We see that temperature fluctuations are large enough to exceed the noise only for large heat fluxes ($\Phi>3$~W). The amplitude of fluctuations increases with the heating flux. The order of magnitude of temperature fluctuations we measure in our setup is $\sigma( T')\sim0.05\text{~K}$, for the maximum heat flux $\Phi=5.8$~W. For smaller heat fluxes ($\Phi<3$~W), the signals are dominated by the electronic noise with a constant apparent standard deviation of 0.02~K. For large heat fluxes ($\Phi>3$~W), we show the calculated power spectral density of the temperature signals on Fig.~\ref{fig:temperature-spectrum} which shows a good fit to a $-5/3$ slope.

\begin{figure}[ht!]
\begin{center}
\includegraphics{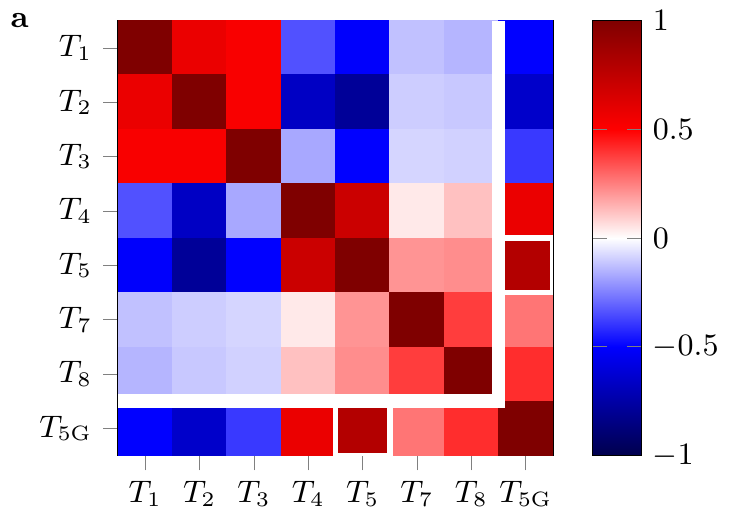}
\includegraphics{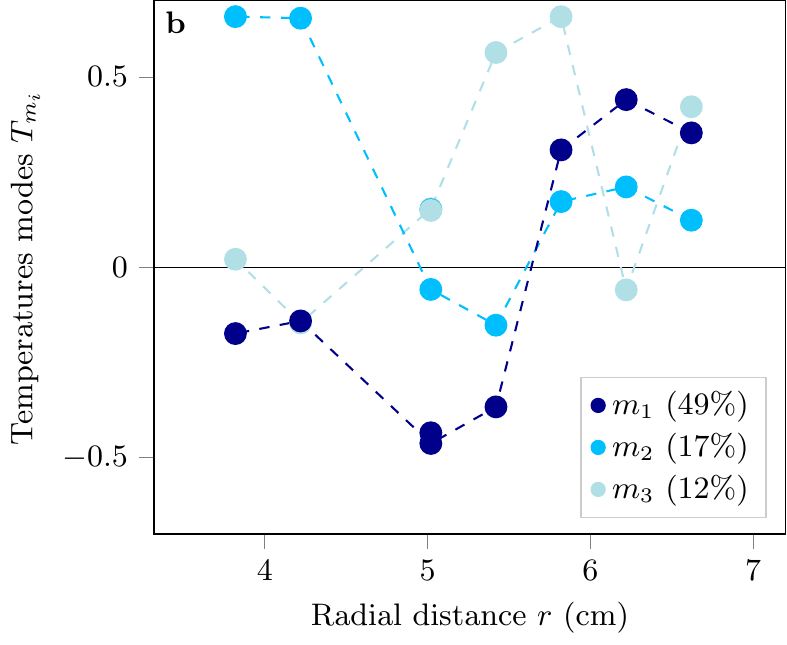}
\caption{\label{fig:correlmatrix} (a) Correlation matrix between temperature fluctuations signal at 7000~rpm and $\Phi=4.0\pm0.2$~W. The signal $T_{5\text{G}}$ corresponds to the thermistor opposite $T_5$. The high-correlation between $T_{5\text{G}}$ and $T_5$ validates the geostrophic hypothesis.(b) Plots of the first three temperatures fluctuation modes inside the cell obtained by a POD analysis. The number in brackets corresponds to the fraction of energy (norm $L^2$) associated with the mode.}
\end{center}
\end{figure}

We cross-correlate fluctuation signals from the different thermistors in the experiment (Fig.~\ref{fig:correlmatrix}(a)). Thermistors can be separated into two groups. The first 3 thermistors close to the heating plate are mutually correlated and the same holds for the farthest thermistors. These two groups of thermistors are anti-correlated. This shows that the flow has a large scale structure inside the cell. This fact is confirmed by a proper orthogonal decomposition (POD) of the signals, providing the main temperature modes in the cell (Fig.~\ref{fig:correlmatrix}(b)). The first mode which represents the majority of the signal energy has the same shape as described above: the first three thermistors evolve together in phase opposition with the others.

In addition, the cross-correlation between the signals from the two thermistors facing each other on Ekman walls, $T_5$ and $T_{5\text{G}}$, is a good way to test the flow geostrophy. There are two opposite effects which influence the geostrophy. The high rotation rate favours a geostrophic flow in the cell whereas convection tends to destroy it by creating a turbulent flow when the heating is strong enough. Since the normalized correlation between these two signals has a high value ($>0.5$) for all the heating fluxes tested, we conclude that the flow in the cell is geostrophic in all our experiments.

\subsection{Pressure measurements}

\begin{figure}
\includegraphics{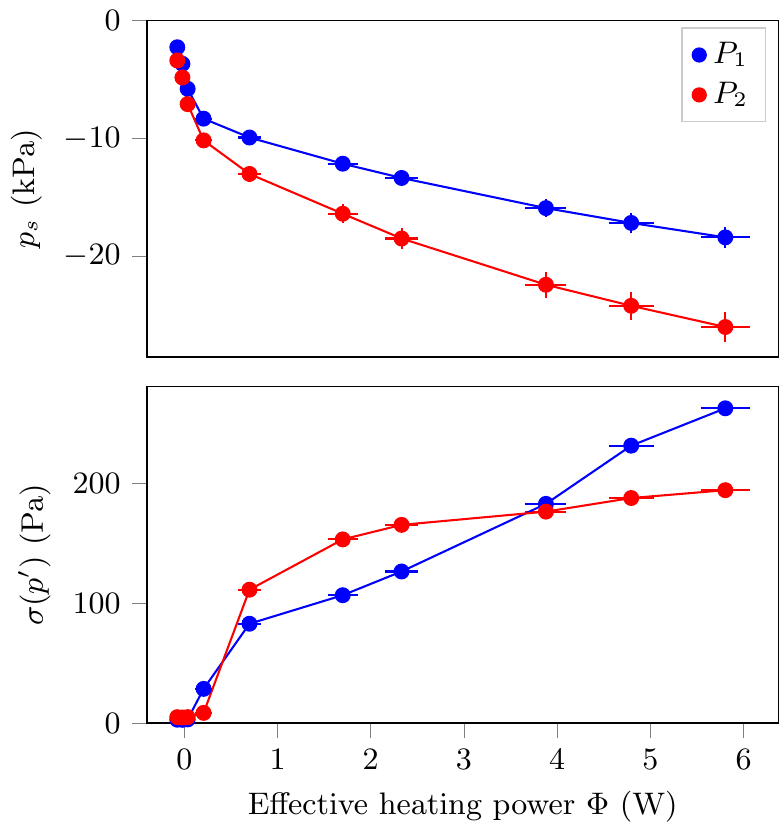}
\caption{\label{fig:STD} Stationary pressure $p_s$ and standard deviation of pressure fluctuations $\sigma(p')$ as a function of the heating power $\Phi$ at 9990 rpm.}
\end{figure}

In a statistically stationary state, similarly to temperatures, the total pressure $p$ is split into three terms 
\begin{equation}
p = p_a + p_s + p',
\end{equation}
where $p_a$ is the adiabatic pressure profile as defined in Eq.~(\ref{eq:gradadiab-P}), $p_s$ is the stationary pressure caused by the large scale stationary flow in the cell. As can be seen on Fig.~\ref{fig:profilT}, pressure signals have a large stationary component that depends only on the heat flux (we will see that it also depends on the rotation rate). At last, $p'$ is the time-dependent pressure fluctuation. The differential pressure probes give us measurements of the dynamical pressure $p-p_a = p_s+p'$. To separate $p_s$ and $p'$ in our analysis, we consider that fluctuations $p'$ corresponds to the part of the signal above 1~Hz. The part below 1~Hz has a constant value during stationary states which corresponds to $p_s$. This decomposition of the pressure leads to consider three terms with very different order of magnitude: $p_a$ is of order of 2~MPa, $p_s$ is of order of 10~kPa and $p'$ is of order of 100~Pa (see Fig.~\ref{fig:STD}).

From the stationary pressure $p_s$, we estimate the order of magnitude of the velocity $v$ in the cell. Assuming geostrophic equilibrium in the fluid, we expect the pressure gradient $-\vect\nabla p_s$ to balance the Coriolis acceleration $2\rho_a\vect\Omega\times\vect v$. With the typical length-scale $L$, we find an estimate for $v$
\begin{equation}\label{eq:estimgeo}
v\sim \frac{p_s}{2 \rho_0 \Omega L} \sim 1\text{~m~s$^{-1}$}.
\end{equation}
This estimate leads to a Rossby number of $\mathrm{Ro}\sim 10^{-2}$, which is consistent with the initial assumption of geostrophic balance. Alternatively, assuming that the pressure gradient is balanced by 3D inertial terms $\rho_a ( {\vect v} \cdot {\vect \nabla} ) {\vect v}$,  we then obtain the following velocity estimate
\begin{equation}
v \sim \sqrt{\frac{p_s}{2 \rho_0 }}  \sim 10\text{~m~s$^{-1}$} , 
\end{equation}
leading to a Rossby number of $0.1$, which seems too small to justify a 3D balance. Moreover, the excellent correlation of the temperature signals of the probes $T_5$ and $T_{5\text{G}}$ gives further credential to the quasi-geostrophic dynamics.
To conclude, the estimate provided by Eq.~(\ref{eq:estimgeo}) is most probably correct.

At this point, we have determined the order of magnitude of the pressure and temperature departures from their hydrostatic and adiabatic profiles: $T_s \sim 1$~K and $p_s \sim 10$~kPa. This can be used to evaluate their relative contribution to the departure of entropy from a uniform value. Gibbs equation $T \mathrm{d} s = c_p \mathrm{d} T - \alpha T/ \rho \mathrm{d} p$ allows us to determine the ratio of the pressure contribution to the temperature contribution in the entropy variations
\begin{equation}
\frac{\alpha T}{\rho c_p} \frac{p_s}{T_s} \simeq \frac{\alpha_s}{\rho_0} \frac{p_s}{T_s}\sim 0.6, 
\end{equation} 
while \citet{anufriev2005} give an estimate $(\alpha T ) \mathcal{D} = 0.06$ for the same ratio, see their equation (2.17). This ratio is crucial to decide whether the so-called anelastic liquid approximation can be used, whereby entropy departures are expressed in terms of temperature departures only. From our experiments it seems that pressure departures are underestimated in \cite{anufriev2005}. They use a balance between buoyancy forces and pressure gradient, to derive pressure departures from hydrostatics as a function of temperature departures from the adiabatic profile (their equation (2.17)a in a different form)
\begin{equation}\label{buoy_press}
p_s \sim \rho_0 \alpha g L T_s .
\end{equation}
This balance does not apply to our data: with $T_s \sim 1$~K, Eq.~(\ref{buoy_press}) leads to $p_s \sim 1000$~Pa whereas the measured value is of order $p_s\sim 10$~kPa, see Fig.~\ref{fig:STD}.

\begin{figure}
\includegraphics{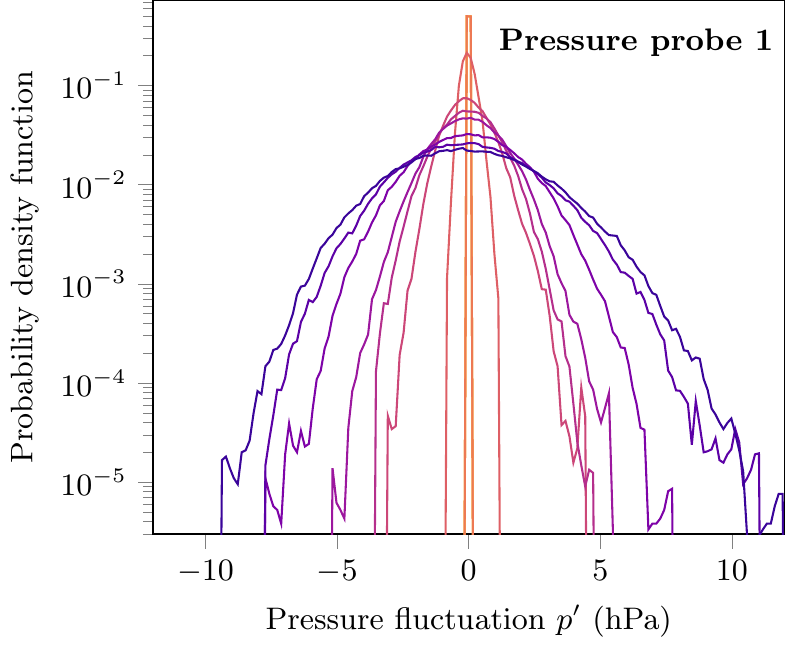}
\includegraphics{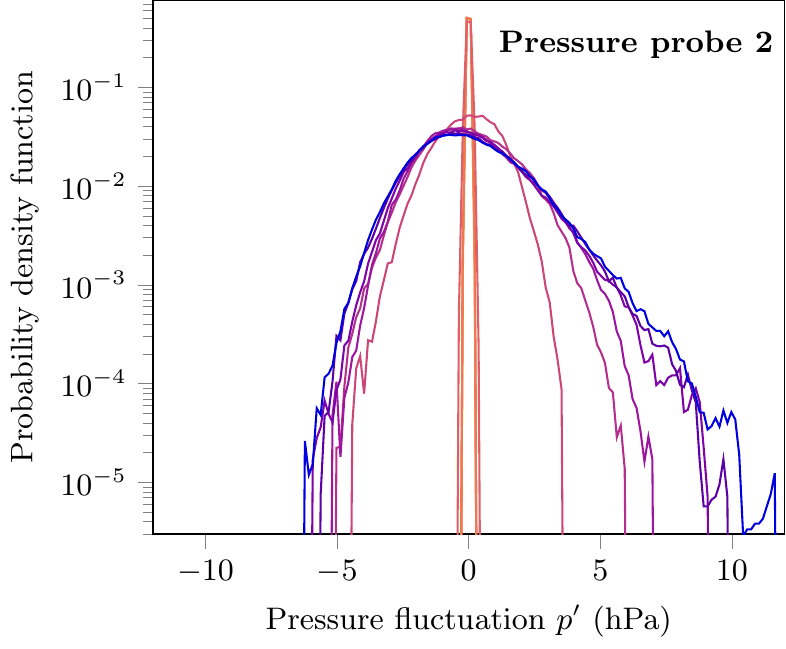}
\caption{\label{fig:PDF} Probability density function of pressure fluctuations for both pressure sensors (pressure probe 1 is the closest to the heating plate) at 9990 rpm for several heating powers. Colors are the same as in Fig.~\ref{fig:gradadiab}(b).
}
\end{figure}

\begin{figure}
\includegraphics{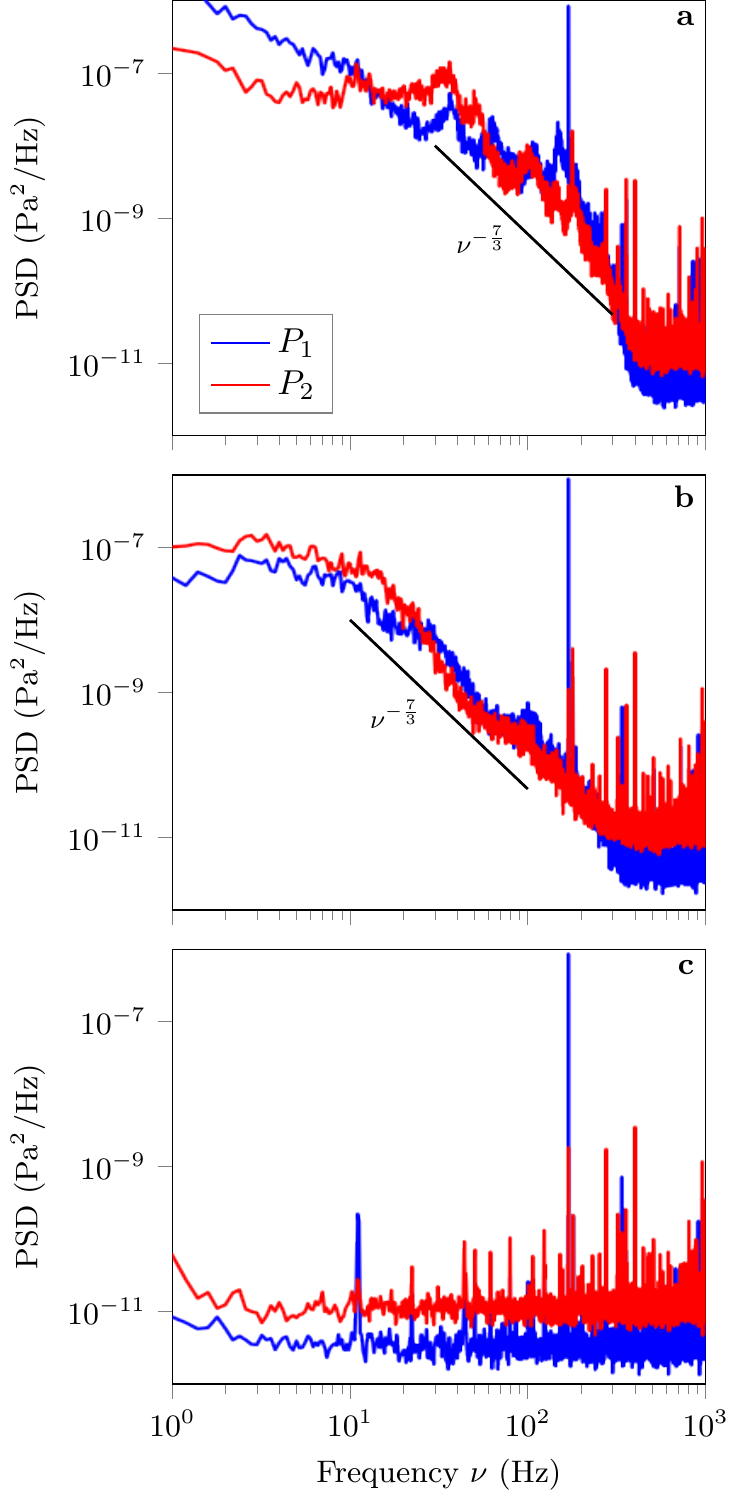}
\caption{\label{fig:pressureFFT} Power spectral density of pressure fluctuations for the same experiment at 9990 rpm (166.5~Hz) and for three different heating fluxes: (a) $\Phi=(4.8 \pm 0.2)$~W, (b) $\Phi=(0.7 \pm 0.1)$~W, \textbf{(c)} $\Phi=0$~W. The black line is a -$7/3$ slope. The peak at 11~Hz corresponds to signal cross-talking with the cycle measurement rate for temperatures.}
\end{figure}

To correctly evaluate the pressure fluctuations $p'$ we eliminate the long-term variations of the signal below 1~Hz and short-term variations above the rotation frequency. At higher frequency, the signal is dominated by peaks of harmonics of the rotation frequency and electronic noise, while the hydrodynamic part of the signal is already nearly at the level of noise measurement, as we will see later on Fig.~\ref{fig:pressureFFT}. The amplitude of the fluctuations depends strongly on the heat dissipated in the heating resistor. This amplitude goes to zero when the convection stops due to a insufficient heating. We estimate the amplitudes of fluctuations by taking the standard deviation of the filtered signal (Fig.~\ref{fig:STD}). We calculate the probability density function (PDF) of pressure fluctuations for each heating power tested. The PDF of the signal from probe 1 is nearly gaussian (Fig.~\ref{fig:PDF}, top), while the PDF of the signal from probe 2 shows a distinct asymmetry with a longer tail on the positive fluctuations side (Fig.~\ref{fig:PDF}, bottom). 

From the temperature and pressure fluctuations, we can test equation (\ref{buoy_press}) originating from the dynamical balance between buoyancy and pressure gradient. At $9990$~rpm and for large heat fluxes, using $\sigma(T') \sim 0.05$~K, equation (\ref{buoy_press}) leads to pressure fluctuations of order $50$~Pa, whereas we measure $\sigma(p') \sim 100$~Pa. Contrary to the stationary departures $p_s$ and $T_s$, pressure and temperature fluctuations $p'$ and $T'$ seem to obey rather well the balance between buoyancy forces and pressure gradient. 

We calculate the power Fourier spectra of $p'$ (Fig.~\ref{fig:pressureFFT}). The spectra are dominated by very localized peaks corresponding to the rotation rate and harmonics. Moreover, there are some other very localized peaks above 200~Hz which are probably created by the electronics and not relevant to our study. If we look at the general behavior of these spectra, there are two kinds of spectra. For low heating fluxes, the spectra have a constant value, corresponding to the noise level of the acquisition. It is characteristic of the piezoelectric sensor we use since a similar spectrum is obtained outside the cell without any load. For higher heating fluxes, the spectra have three parts. The first part, below 13~Hz, has an approximatively constant value. The second one, between 13~Hz and 20-40~Hz, decreases with a slope close to $-7/3$ (in log-log coordinates) which is the expected power law for pressure fluctuations in homogeneous turbulence \cite{batchelor1953}. And the third part, above 40~Hz, is constant and corresponds to the noise level. 

The threshold between these two kinds of spectra is the same as the threshold evaluated in part \ref{ssec:AdiabaticGradient} with the temperature profile. When the heat flux value is too low, there is no convection and we only measure the sensor response. For higher fluxes, the convection is established and we measure the convection spectrum.

\subsection{Convection pattern and heat flux}

Using the results obtained above, we can make an attempt to characterize the structure of the quasi-geostrophic convective flow in our experiment. The pressure signals indicate that the stationary part of the convective flow is much more vigorous than its fluctuations, by a factor $100$. We can get an idea of the magnitude of stationary temperature deviations from the adiabat when looking at Fig.~\ref{fig:gradadiab}(b) and \ref{fig:doublefit}(b). As we already pointed out, the temperature profile is very close to the theoretical adiabatic profile when the heat flux is just exceeding the heat losses. Increasing the heat flux further causes the temperature measurements to depart from the adiabatic profile. The amplitude of the deviations is of the order of 1~K for the maximum heat flux (about 5~W).  We believe that these deviations are linked to the stationary part of the convective flow: this is indeed in accordance with the estimate for the convective heat flux. The convective heat flux is estimated as 
\begin{equation}
\Phi \sim \rho_0 c_p S v \delta T. \label{eq:estimflux}
\end{equation}
For the highest rotation rate (9990~rpm) and highest heat flux (about 5~W), we have estimated $v \sim 1$~m~s$^{-1}$, and $\delta T \sim 1$~K. With a cross-section area $S=530$~mm$^2$, density $\rho_0 \sim 100$~kg~m$^{-3}$ and heat capacity $c_p=204$~J~K$^{-1}$kg$^{-1}$, the estimate Eq.~(\ref{eq:estimflux}) gives $\Phi \sim 13~W$, which is of the same order of magnitude as the actual heat flux (about 5~W), given that we do not have sufficient spatial coverage to extract the shape of that stationary flow and associated thermal signature. In any case, the velocity fluctuations (100 times smaller than the stationary flow) together with the temperature fluctuations (less than $0.05$~K, see Fig.~\ref{fig:profilT}) would by no means generate a convective heat flux close to the imposed heat flux. The large-scale stationary flow and its small fluctuations are also in good agreement with the POD analysis of temperature fluctuations, showing a large-scale pattern (see Fig.~\ref{fig:correlmatrix} (b)). 

The global picture of the convective flow is that of a two dimensional turbulent flow in which the inverse energy cascade has given rise to an intense stationary large-scale flow dominating small-scale fluctuations. This process is also known as the condensation of small scale vortices in a large-scale steady flow and was observed for two-dimensional flows due to rotation, stratification, or the effect of an imposed magnetic field on an electrically conducting fluid \cite{hossain1994,paret1997,sommeria1986}. 

	Within the frame of this global picture, we can re-examine some of the results presented above. Concerning the spectrum of temperature fluctuations on Fig.~\ref{fig:temperature-spectrum}, the signal above 2 or 3~Hz is dominated by measurement noise. With a global flow of 1~m~s$^{-1}$ and a cavity height of $0.039$~m, we expect that the turn-over frequency of large vortices is of order 20~Hz. Hence the temperature spectrum corresponds to very low frequencies which are probably representative of the long-term evolution of the main large-scale circulation and does  not give information on the small-scale structure of turbulence. On the contrary, pressure spectra shown on Fig.~\ref{fig:pressureFFT} allow us to determine the frequency of the large flow structures, corresponding to the kink point between a flat part at low frequency to a steep part at higher frequency. That steep part contains probably information on the small scale turbulence of the flow. The low frequency nature of the temperature fluctuations is also probably the reason why the POD analysis, see Fig.~\ref{fig:correlmatrix}, produces large scale modes which correspond to the slow evolution modes of the large scale circulation.

\subsection{Initial transient}

\begin{figure}
\includegraphics{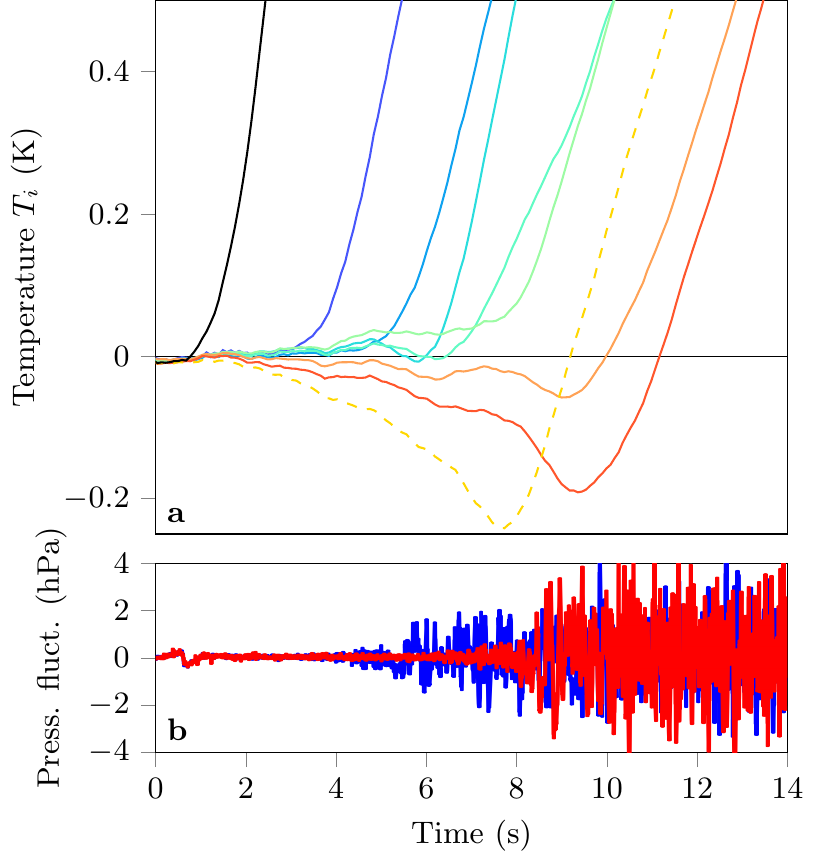}
\includegraphics{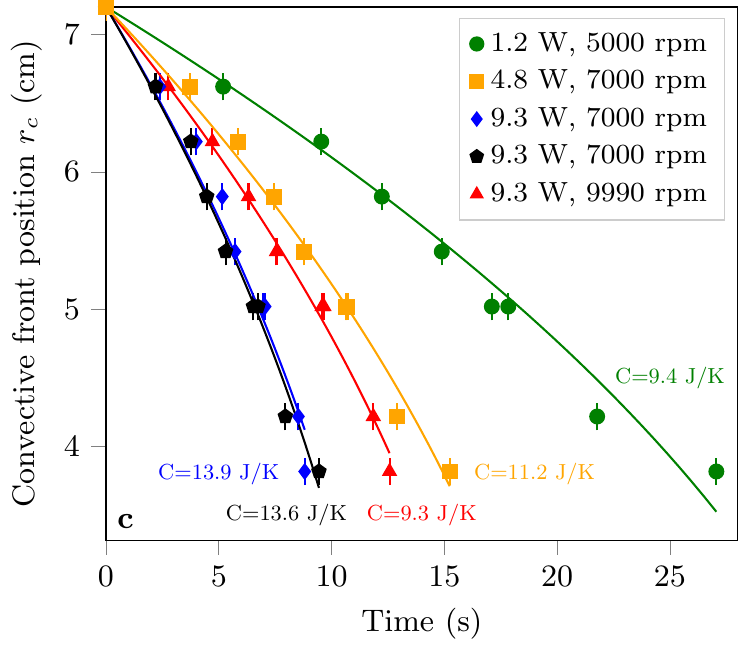}
\caption{\label{fig:front} (a-b) Temperature signals and pressure fluctuations versus time since the beginning of heating. The color code is the same as in Fig.~\ref{fig:profilT}. The dashed yellow line corresponds to $T_{5\text{G}}$. (c) Position of the convective front $r_c$ versus time. Solid lines are theoretical evolutions from Eq.~(\ref{eq:front}) with the adjusted value of $C$.}
\end{figure}

	The temperature signals allow us to study how convection is established in the cell at the beginning of an experiment. The initial isothermal state corresponds to a strongly stratified stable configuration (entropy increases with height). When we start to heat from below, we see the temperature signals starting to increase one after another (Fig.~\ref{fig:front}(a)) indicating that a convective region develops. If we plot the starting time of each signal versus the position of the corresponding thermistor, we track the front of convection in the cell. We can also track the front of convection by looking at the beginning of pressure fluctuations on both pressure probes. This corresponds to a case of penetrative convection, as a growing convective region erodes an initially stagnant region. 

	We propose here a simple model to describe the front propagation. We assume that all the setup (xenon and walls) is initially at a constant temperature $T_i$. At time t=0, the heating starts at a constant power flux $\Phi_T$. The bottom plate, where the front of convection starts, is at $r=r_{max}$. For $t>0$, the position of the front of convection is denoted by $r_c(t)$.  During the onset of convection, $r_c(t)$ goes from $r_{max}$ to $r_{min}$. At a given time $t$, the temperature profile in the cell has to follow the adiabatic profile in the convective area and stays at $T_i$ outside. Thus, it has the form
\begin{equation}
T(r,t) = \left\{\begin{array}{lc} 
T_i + \frac{\Omega^2}{2c_p} (r^2 - r_c(t)^2) & \text{for } r\geq r_c(t),\\
T_i & \text{for } r< r_c(t).
\end{array}\right.
\end{equation}

	The total heat capacity of xenon is negligible compared to that of the walls. Thus, the majority of energy is used to heat the walls. A simple model is to consider that the internal energy of the walls is proportional to the bottom plate temperature $T(r_{max},t)$. The proportionality constant $C$ is an effective heat capacity. The energy balance is then
	\begin{equation}
	\Phi_T t = C (T(r_{max},t)-T_i).
	\end{equation}
	By inverting this equation, we find
	\begin{equation}\label{eq:front}
	r_c(t) = r_{max}\sqrt{1-\frac{2 c_p \Phi_T t}{C \Omega^2 r_{max}^2}}.
	\end{equation}

	The shape given by this expression is in accordance with our data where we see an acceleration of the front of convection. We adjust on Fig.~\ref{fig:front}(c) the value of $C$ in Eq.~(\ref{eq:front}) to obtain the best fit of our data points. According to the experimental data, we find $C$ between 9~J/K and 14~J/K. However with the same conditions, we find the same value of $C$. The results show that $C$ depends on the heat flux $\Phi_T$ and the rotation rate $\Omega$. These values of $C$ have the same order of magnitude as the heat capacity of the whole setup evaluated in section \ref{ssec:heatlosses}, although this simplified model ignores convection (and rotation effects). Nevertheless, we see on the experimental temperature curve that there is a small cooling of some tenth of Kelvin just before the front of convection reaches the thermistor and the temperature increase significantly. This phenomenon is more and more important during the front propagation and is even visible on the thermistor on the opposite wall. This is typically a compressible effect and could never be envisaged in the Boussinesq  approximation. We temporarily observe temperature levels lower than the initial uniform temperature, while the bottom boundary gets heated: this is a manifestation of compressible penetrative convection. A fluid parcel initially at rest can be cooled adiabatically when it is suddenly entrained into convection as the convective region reaches its position. 

\section{Conclusion}
\label{conclusion}


In the reported experiments, we show that compressible convection can be studied in the laboratory. A substantial adiabatic temperature gradient (about $3.5$~K~cm$^{-1}$) has been measured. The Nusselt-Rayleigh relationship, once the adiabatic gradient has been taken into account, follows the expected $1/3$ power law. Adiabatic decompression effects are observed during the transient heating of the fluid from an initial isothermal state. However, the dimensionless dissipation number reaches only modest values, around $0.06$. This probably means that compressible effects are moderate and that the experiments are within what we might call a 'Jeffreys' regime: \cite{jeffreys1930} has shown that the stability criterion derived by Rayleigh for the onset of convection was still applicable, provided the superadiabatic temperature difference is considered instead of the total imposed temperature difference. It can indeed be seen \cite{alboussiere2017} that Jeffreys' result holds for a dissipation number as small as $0.06$. Away from the stability threshold, numerical results from the literature also suggest that compressible results are very close to those obtained using the Boussinesq approximation \cite{verhoeven2015, tilgner2011,curbelo2018sub}. Our experiments nevertheless open the way to further attempts and further progress into experimental compressible convection. We are still working on the possibility to make measurements with twice the maximal rotation rate reported here, {\it i.e.} 20000~rpm. This would increase gravity levels by a factor four, leading to a dissipation number around $0.24$. Next, using another centrifuge or other dedicated experiments, it is conceivable to obtain dissipation numbers of order $1$ or more, and even to run compressible convection experiments with other fluids. 

During our experiments, we have shown that temperature measurements were possible in a hostile environment of apparent high gravity. We have been using small thermistors probes for two reasons: first, small parts have a relatively larger resistance to stress and can sustain better high gravity. Secondly, we wanted to measure fast temperature fluctuations in a gas with small heat capacity and were looking for a probe with a small thermal diffusion time and small heat capacity. In addition, we have also successfully used differential pressure probes ($100$~Pa), under large pressure level ($3\ \times \ 10^6$~Pa) and under a large gravity ($7\ \times \ 10^4$~m~s$^{-2}$). Those signals were processed using onboard home-designed electronics, essentially multiplexing signals, which worked satisfactorily up to 9990~rpm, but did not allow us to obtain results at higher rotation rates, due to the failure of an electronic component. It is difficult to assess whether electronic components will sustain high gravity levels based on their datasheet. The method is simply to go for small components and test them.

Since we aim for a large apparent gravity level, and since we are using a low-viscosity fluid, our experiments have large values for the (superadiabatic) Rayleigh number and small values for the Ekman number. In fact, Coriolis effects are so large that all our experimental runs obey a quasi-geostrophic dynamics. This is a difficulty as we would like to unravel the role of compressibility and that of geostrophy: for instance, is the hysteresis behaviour between branches of a large heat flux and a small heat flux due to geostrophy, or compressibility, or both?
Concerning the value of the superadiabatic Rayleigh number, one could think that it would be made as small as desired by just reducing the imposed heat flux (and associated superadiabatic temperature difference). However, we have shown that our cavity has a significant amount of heat losses, despite the use of an excellent insulating material (aerogel). This imposes a minimum level to the heat flux that can effectively be studied. In our case, this limits our range of superadiabatic Rayleigh numbers to $5\ \times \ 10^{11}$ from below. As a corollary, the study of the onset of convection seems hopeless in our setup and similar ones. \citet{busse2014} had anticipated that some aspect of compressible convection could be studied from the oscillatory or stationary character of the first unstable mode of convection, but large difficulties should be expected to observe this first mode.

Which of our measurements are due to compressibility effects and which are not?
Typically, the temperature gradient along the direction of the apparent gravity,
{\it i.e.} the adiabatic gradient, is certainly due to compressibility. Secondly, 
 the negative overshoot of temperature during the initial transient after heating the bottom plate can only be understood within the frame of compressible convection. Concerning the heat flux results, we have obviously removed the effect of the adiabatic gradient in our estimate of the effective driving temperature difference of convection. After this is done (and this is an important aspect of compressible convection), we obtain a classical (incompressible) $\Nu-\mathrm{Ra}$ relationship. Stated otherwise, the isolated bottom boundary layer behaves like an incompressible convection boundary layer in terms of the heat flux going through it as a function of the temperature difference across it. The whole initial transient is also due to compressibility: the isothermal initial state corresponds to a strongly stably stratified situation (in terms of entropy), whereas this initial state would be marginally stable in the incompressible case. Geostrophy and the amplitude of pressure measurements are probably due to Coriolis forces and independent of compressibility. The modes of temperature fluctuations (Fig.~\ref{fig:correlmatrix}) are also probably mostly due to Coriolis effects on convection, however we do not know at this stage whether compressibility has an effect on them. 

Despite the difficulties, our experiment shows that it is indeed possible to study compressible convection in the laboratory. This should provide future benchmarks, and unravel new phenomena to be compared to compressible convection models and to natural phenomena observed in stars and planets. \\

\begin{acknowledgments}
The authors are grateful to the LABEX Lyon Institute of Origins (ANR-10-LABX-0066) of the Université de Lyon for its financial support within the program "Investissements d'Avenir" (ANR-11-IDEX-0007) of the French government operated by the National Research Agency (ANR). Thanks are due to the program PNP of INSU for its financial support. We also thank Christophe Gal\'ea and CDB Electronique for helping us to make the electronic card.
\end{acknowledgments}

\bibliography{bib_papier}

\end{document}